\documentclass[12pt]{article}
\usepackage{epsfig,cite}
\usepackage {amsmath}
\usepackage{amssymb}
\include{epsf}
\newcount\timecount
\newcount\hours \newcount\minutes  \newcount\temp \newcount\pmhours
\hours = \time
\divide\hours by 60
\temp = \hours
\multiply\temp by 60
\minutes = \time
\advance\minutes by -\temp
\def\hour{\the\hours}
\def\minute{\ifnum\minutes<10 0\the\minutes
            \else\the\minutes\fi}
\def\clock{
\ifnum\hours=0 12:\minute\ AM
\else\ifnum\hours<12 \hour:\minute\ AM
      \else\ifnum\hours=12 12:\minute\ PM
            \else\ifnum\hours>12
                 \pmhours=\hours
                 \advance\pmhours by -12
                 \the\pmhours:\minute\ PM
                 \fi
            \fi
      \fi
\fi
}

\def\monthname{\relax\ifcase\month 0/\or January\or February\or
   March\or April\or May\or June\or July\or August\or September\or
   October\or November\or December\else\number\month/\fi}

\def\bold#1{\setbox0=\hbox{$#1$}%
     \kern-.025em\copy0\kern-\wd0
     \kern.05em\copy0\kern-\wd0
     \kern-.025em\raise.0433em\box0 }

\def\beq{\begin{equation}}
\def\eeq{\end{equation}}
\def\bear{\begin{eqnarray}}
\def\eear{\end{eqnarray}}

\def\ga{\mathrel{\raise.3ex\hbox{$>$\kern-.75em\lower1ex\hbox{$\sim$}}}}
\def\la{\mathrel{\raise.3ex\hbox{$<$\kern-.75em\lower1ex\hbox{$\sim$}}}}
\def\gev{{\rm \, Ge\kern-0.125em V}}
\def\tev{{\rm \, Te\kern-0.125em V}}
\def\gyr{{\rm \, G\kern-0.125em yr}}

\def\slash#1{\rlap{\hbox{$\mskip 1 mu /$}}#1}%

\def\gappeq{\mathrel{\rlap {\raise.5ex\hbox{$>$}}
{\lower.5ex\hbox{$\sim$}}}}
\def\lappeq{\mathrel{\rlap{\raise.5ex\hbox{$<$}}
{\lower.5ex\hbox{$\sim$}}}}
\def\Toprel#1\over#2{\mathrel{\mathop{#2}\limits^{#1}}}

\def\stop{\widetilde t}
\def\sbot{\widetilde b}

\def\sM{{\widetilde{\cal M}}}
\def\ts{\tilde{t}}
\def\bal{\begin{array}}
\def\eal{\end{array}}

\def\beqn{\begin{eqnarray}}
\def\eeqn{\end{eqnarray}}

\def\mchi{m_{\tilde \chi}}

\def\m12{m_{1\!/2}}

\def\mstop{m_{\tilde t_1}}

\def\mstau{m_{\tilde{\ell}_1}}

\def\ts{\tilde{t}}
\def\bs{\tilde{b}}

\def\mstau{m_{\tilde{\tau}}}
\def\mgrav{m_{\widetilde G}}

\def\mchi{m_{\chi}}

\def\bea{\begin{eqnarray}}
\def\eea{\end{eqnarray}}

\begin{document}
\begin{titlepage}
\pagestyle{empty}
\baselineskip=21pt
\renewcommand{\thefootnote}{\fnsymbol{footnote}}
\rightline{\tt hep-ph/0701229}
\rightline{CERN-PH-TH/2007-003}
\rightline{UMN--TH--2536/07}
\rightline{FTPI--MINN--07/03}
\vskip 0.2in
\begin{center}
{\large{\bf On the Feasibility of a Stop NLSP in Gravitino Dark Matter Scenarios}}
\end{center}
\begin{center}
\vskip 0.2in
{\bf J. L. Diaz-Cruz}$^1$\footnote{
On sabbatical leave at: Facultad de Ciencias, Universidad de
Colima, M\'{e}xico.}, {\bf John~Ellis}$^2$, {\bf Keith~A.~Olive}$^{3}$
and {\bf Yudi Santoso}$^{4}$
\vskip 0.1in

{\it
$^1${Facultad de Ciencias F\'{i}sico-Matem\'{a}ticas, BUAP \\
Apdo. Postal 1364, C.P.72000 Puebla, Pue, M\'{e}xico}\\
$^2${TH Division, CERN, Geneva, Switzerland}\\
$^3${William I. Fine Theoretical Physics Institute, \\
University of Minnesota, Minneapolis, MN 55455, USA}\\
$^4${Department of Physics and Astronomy, University of Victoria,\\
 Victoria, BC, V8P 1A1, Canada}}

\vskip 0.2in
{\bf Abstract}
\end{center}
\baselineskip=18pt \noindent

We analyze the possibility that the lighter stop ${\tilde t_1}$ could be the
next-to-lightest supersymmetric particle (NLSP) in models where the gravitino
is the lightest supersymmetric particle (LSP). We do not find any possibility
for a stop NLSP in the constrained MSSM with universal input soft
supersymmetry-breaking masses at the GUT scale (CMSSM), but do find small allowed
regions in models with non-universal Higgs masses (NUHM). We discuss 
the cosmological evolution of stop hadrons. Most ${\tilde t_1}qq$
`sbaryons' and  the corresponding `antisbaryons' annihilate with conventional
antibaryons and  baryons into ${\tilde t_1}{\bar q}$ `mesinos' and the
corresponding `antimesinos', respectively, shortly after the quark-hadron
transition in the early Universe, and most mesinos and antimesinos
subsequently annihilate. As a result, insufficient metastable charged stop 
hadrons survive to alter Big Bang nucleosynthesis.

\bigskip
\leftline{CERN-PH-TH/2007-003}
\leftline{January 2007}
\bigskip
\bigskip
\end{titlepage}
\baselineskip=18pt
\renewcommand{\thefootnote}{\arabic{footnote}}

\section{Introduction}

In many supersymmetric models there is a multiplicatively-conserved quantum
number, $R$ parity, that guarantees the stability of the lightest
supersymmetric particle (LSP). In order to avoid the LSP binding to ordinary
matter, it is usually assumed to have neither strong nor electric
charge~\cite{EHNOS}. Candidates for the LSP in the minimal supersymmetric
extension of the Standard Model (MSSM) with gravity include sneutrinos, the
lightest neutralino $\chi$ and the gravitino ${\widetilde G}$. Light sneutrinos
were  excluded by searches for invisible $Z$ decays at LEP, and heavier stable
sneutrinos would have been found in direct searches for the scattering of
astrophysical dark matter particles on ordinary matter~\cite{FOS94}. Thus, most
attention has focused on the neutralino and the gravitino. Overlooked to some
extent, a gravitino LSP is in fact quite generic even in models based on
minimal supergravity (mSUGRA) \cite{vcmssm}.

In the case of a gravitino
LSP~\cite{gdm,FengGDM,FengGDM2,FengGDM3,otherGDM2,otherGDM3,steffan}, the
next-to-lightest supersymmetric particle (NLSP) has a long lifetime, decaying
with gravitational-strength interactions if supersymmetry breaking is mediated
by supergravity. The question of the identity of the NLSP then becomes
important. One generic possibility is that the NLSP is the lightest neutralino,
in which case the long-lived $\chi$ would probably decay unseen, mainly via
$\chi \to {\widetilde G} + \gamma$,  without being stopped beforehand \cite{gdm,FengGDM}. 
Another generic possibility is the lightest charged slepton, probably the lighter stau
${\tilde \tau_1}$ in the MSSM with universal scalar soft supersymmetry-breaking
masses (the CMSSM)~\cite{gdm,FengGDM2,FengGDM3,otherGDM3}. 
This leads to scenarios with a
metastable charged sparticle that  would have dramatic signatures at
colliders~\cite{Bench3,FengGDM2,Are,Feng+Smith,Nojiri} and could affect
drastically the
cosmological abundances  of light elements.

Electromagnetic showers from the decay products of metastable particles can
alter the abundances of light elements by photo-dissociation and subsequent
secondary reactions~\cite{othercosm,CEFO,EOV}.  Moreover, hadronic showers can alter the
amounts of baryons involved in the Big-Bang nucleosynthesis (BBN) processes  if
the lifetime $\lappeq 10^{6}$~s~\cite{KKM}. However, a more significant effect
can occur in the case of a negatively-charged particle, which can form an
electromagnetic  bound state with a nucleus, and influence the BBN processes by
lowering the Coulomb barrier for nuclear fusion~\cite{Maxim,otherBoundS} (a
catalytic effect). This has been studied within the GDM in the case of a stau
NLSP in some CMSSM and mSUGRA scenarios~\cite{CEFOS}. 

However, there are also other possible candidates for the NLSP, such as some
sneutrino~\cite{FengGDM3,KKKM} or  squark species. Among the different squark
species, a generic candidate for the lightest is the lighter stop ${\tilde
t_1}$~\cite{ER}, which would have interesting implications for cosmology~\cite{BDD,stopco}, 
although there are other possibilities. In this paper we
study the feasibility of scenarios with a gravitino LSP and a stop NLSP. Thus, 
we search for  regions of the MSSM parameter space where the $\stop_1$ is
lighter than the supersymmetric  partners of all the other Standard Model
particles, including the neutralino $\chi$. A previous  study showed that this
is possible for large values of the soft trilinear supersymmetry-breaking
parameter $A_0$, and the regions of the CMSSM parameter space where  this
happens have been  delineated~\cite{stopco}. In traditional scenarios with
conserved  $R$ parity and a heavy gravitino, these regions would have been
discarded because they have a charged and coloured stable particle.  However,
if  the gravitino $\widetilde{G}$ is the LSP and therefore constitutes the dark
matter, one should explore whether some parts of these regions  might
survive.   

There are several experimental and cosmological constraints on such a stop NLSP
scenario that must be taken into account. As discussed in more detail below,
the lifetime of the ${\tilde t_1}$ may be (very) long, in which case the
relevant collider limits are those on (apparently) stable charged particles. We
interpret the limits available from the Tevatron collider as implying that
$m_{\tilde t_1} > 220$~GeV~\cite{Tevatron}~\footnote[1]{The LHC will probably be sensitive to a
metastable ${\tilde t_1}$ that is an order of magnitude heavier.}. We find no
regions of the CMSSM parameter space compatible with this and other
experimental constraints on ${\tilde t_1}$ NLSP scenarios. However, when we
relax the CMSSM universality assumptions by considering non-universal soft
supersymmetry-breaking masses for the Higgs fields (NUHM models), we do find
limited regions of parameter space with a ${\tilde t_1}$ NLSP. Typical allowed
values of the NUHM parameters are $m_{1/2} \sim 600$~GeV, $m_0 \sim 500$~GeV,
$A_0 \sim 2100$~GeV, $\mu \sim 750$~GeV, $m_A \sim 1400$~GeV and
$\tan \beta \sim 10$.

We then consider the cosmological constraints on such cases. As we show, the
density of ${\tilde t_1}$ sparticles and antisparticles after cosmological
freeze-out at a temperature of several GeV is strongly suppressed
by strong couplings in the annihilation processes.  Subsequently, at
the quark-hadron transition these stops would have combined with
quarks into ${\tilde t_1}qq$ `sbaryons' and ${\tilde t_1}{\bar q}$ `mesinos'
and the corresponding antiparticles. The late decays of these stop hadrons
could have affected the light-element abundances obtained from Big-Bang
nucleosynthesis, and negatively-charged antisbaryons and antimesinos could have
had dramatic bound-state effects. However, we argue that the great majority of
the stop antisbaryons would have annihilated with conventional baryons to make
stop antimesinos, and that most mesinos and antimesinos would subsequently have 
annihilated~\cite{luty}. 
Any negatively-charged antimesinos would have decayed
(relatively) rapidly into neutral mesinos. These would have been (almost) the
only metastable ${\tilde t_1}$ relic particles, and would be relatively
innocuous, despite their long lifetimes, because they would not have important 
bound-state effects. Because of the low density of ${\tilde t_1}$ after
freeze-out following coannihilation and the subsequent cosmological evolution, 
this limited region of stop NLSP scenarios
within the NUHM framework seems to be viable. We conclude our paper with a
brief discussion how such a scenario could be probed experimentally.

\section{Stop Properties}

\subsection{Stop Masses, Mixing and Couplings}

We start by giving some  important formulae and making some crucial definitions.
The (2x2) stop mass matrix may be written as:
\beq
\sM^2_{\ts} =\left\lgroup
         \bal{ll}
          M_{LL}^2         &  M_{LR}^2\\[1.5mm]
          M_{LR}^{2\,\dag}   &  M_{RR}^2
         \eal
         \right\rgroup,
\label{eq:MU6x6}
\eeq
where the entries take the forms:
\beq
\bal{ll}
M_{LL}^2 &= M_{\ts_L}^2+m_t^2+\frac{1}{6}\cos2\beta \,(4m_W^2-m_Z^2)\,, \\[2mm]
M_{RR}^2 &= M_{\ts_R}^2+m_t^2+\frac{2}{3}\cos2\beta\sin^2\theta_W\, m_Z^2\,, \\[1mm]
M_{LR}^2 &= -m_t (A_t + \mu \, \cot \beta) \equiv - m_t X_t\,.
\eal 
\eeq
The corresponding mass eigenvalues are given by:
\beq
m^2_{\ts_1}=m^2_t + \frac{1}{2}(M_{\ts_L}^2+  M_{\ts_R}^2)+
\frac{1}{4}m^2_Z \cos 2\beta-\frac{\Delta}{2}   ,   
\eeq
and
\beq
m^2_{\ts_2}= m^2_t + \frac{1}{2}(M_{\ts_L}^2+  M_{\ts_R}^2)+
\frac{1}{4}m^2_Z \cos 2\beta+\frac{\Delta}{2}    ,            
\eeq
where
$\Delta^2= \left( M_{\ts_L}^2 -  M_{\ts_R}^2 + \frac{1}{6} \cos 2\beta (8
m^2_W-5m^2_Z) \right)^2 + 4\, m_t^2 |A_t + \mu \cot \beta |^2$.
The mixing angle $\theta_{\ts}$ between the weak basis $(\ts_L,\ts_R)$ and the mass
eigenstates  $(\ts_1,\ts_2)$, is given by
$\tan \theta_{\ts}= (m^2_{\ts_1}-M^2_{LL})/|M^2_{LR}|$.
It is clear that
obtaining a very light stop requires  a
very large value for the trilinear soft supersymmetry-breaking parameter~\cite{stopco}.

The interactions of the left and right antistops $\ts^*_L$ and $\ts^*_R$
with the gravitino field ${\bar{\Psi}}_{\mu}$ and the top quark are given 
by~\cite{moroi}~\footnote{Note, however, that there is a typographical error in Eq.(4.31) of
Ref.~\cite{moroi}: $(\slash{p} - m_{3/2})$ should be $(\slash{p} + m_{3/2})$.}:
\beq
{\cal {L}} = -\frac{1}{\sqrt{2} M} [ {\bar{\Psi}}_\mu  \gamma^\nu \gamma^\mu P_R
\, t \, \partial_\nu  \ts^{*}_R +
              {\bar{\Psi}}_\mu \gamma^\nu \gamma^\mu P_L \, t \, \partial_\nu  \ts^{*}_L  ],
\eeq
where the reduced Planck mass is given by
 $M=M_{pl}/\sqrt{8\pi}$, with $M_{pl}=1.2 \times 10^{19}$ GeV.
The interaction lagrangian for ${\ts}_{1,2}$ is then:
\beq
{\cal {L}} =  -\frac{1}{{\sqrt{2}} M} [ {\bar{\Psi}}_\mu  \gamma^\nu \gamma^\mu 
                                ( \sin {\theta}_{\ts} P_R + \cos {\theta}_{\ts} P_L ) 
  \, t \, \partial_\nu  {\ts}^{*}_1 +
                         {\bar{\Psi}}_\mu  \gamma^\nu \gamma^\mu 
                                  (\cos {\theta}_{\ts} P_R - \sin {\theta}_{\ts} P_L ) 
  \,  t \, \partial_\nu  {\ts}^{*}_2  ].
\eeq
The corresponding Feynman rule for the vertex is: 
\beq
 {\ts}^{*}_1 (p) {\bar{\Psi}}^\mu \, t \to  -\frac{1}{\sqrt{2} M}  \, \gamma^{\mu}
 \slash{p} \, ( \sin {\theta} _{\ts} \, P_R + \cos {\theta}_{\ts} \, P_L )  .        
\eeq
Similarly, the Feynman rule for the chargino-gravitino-$W$ vertex is:
\beq
{\chi}^{-}_{i} {\bar{\Psi}}^\mu  {W^\nu}^-(k) \to   -\frac{m_W}{M} \gamma^\nu \gamma^\mu   
                                        ( A_{Li} P_R + A_{Ri} P_L )          
\eeq
Here 
$ A_{Li} =U^\ast_{i2} \cos\beta$, and $A_{Ri}=V^\ast_{i2} \sin\beta$, 
where $V$ and $U$ are the matrices that diagonalize the chargino mass matrix.

\subsection{Stop Decay Modes and Lifetime}

There are several  possible scenarios for stop decay,
depending on the mass difference between the stop NLSP and the gravitino LSP
$\Delta m \equiv \mstop - m_{\widetilde G}$, anticipating that a stop NLSP must
have $\mstop > m_t$ from the direct search bound. 

\begin{enumerate}

\item Case 1: $\Delta m > m_t$, i.e. small $m_{\widetilde G} \lappeq
\mstop - m_t$. In this case, the stop can decay directly into a top 
quark and a gravitino, and the rate for this dominant decay is
\bea
\Gamma &=& \frac{1}{192 \pi} \frac{1}{M_{\rm Pl}^2 m_{\widetilde G}^2 \mstop^3} 
\left[ 4\left( \mstop^2 - m_{\widetilde G}^2 - m_t^2 \right) 
+ 20 \, \sin \theta_{\tilde t} \, \cos \theta_{\tilde t} \, m_t \, m_{\widetilde G} 
\right]  \nonumber \\
&& \times \left[ ( \mstop^2 + m_{\widetilde G}^2 - m_t^2)^2 - 4 \mstop^2 m_{\widetilde G}^2
\right] \left[ ( \mstop^2 + m_t^2 - m_{\widetilde G}^2 )^2 - 4 \mstop^2 m_t^2
\right]^{1/2} .
\label{2bodylife}
\eea
This decay rate is similar to that for stau decay into tau plus gravitino~\cite{gdm}, but
in this case $m_t$ cannot be neglected. Previous results are reproduced in the
limits $m_t \to 0$ and $\theta_{\tilde t} \rightarrow 0$.

We show in Fig.~\ref{fig:2body} some typical numerical results for
$\mstop = 200, 300, 400$ and 500~GeV (from top to bottom),
$m_{\widetilde G} < \mstop - m_t$ and $m_t = 171.4$~GeV~\cite{mt171p4}, 
for both zero stop mixing  (red solid line) and maximal mixing (blue dashed line)
We see that the stop lifetime is relatively insensitive to the stop mixing angle 
$\theta_{\tilde t}$~\footnote{Typical
values in the allowed stop NLSP region in the NUHM are $\theta_{\tilde t} \sim 1.3$.},
but depends sensitively on the sparticle masses, and ranges between
$10^3$ and $10^9$~s. Clearly, this is extremely long compared with the QCD hadronization
time-scale, so that the stop NLSP (unlike the $t$ quark) forms metastable hadrons, whose
spectroscopy and phenomenology we consider below. Moreover, this lifetime range is also
very long on the typical time-scales of collider experiments, which must therefore
consider how to search for these stop hadrons. Indeed, the stop lifetime fits into the range
where the cosmological effects considered in~\cite{CEFO,KKM,Maxim,CEFOS} become
important.

\begin{figure}
\begin{center}
\mbox{\epsfig{file=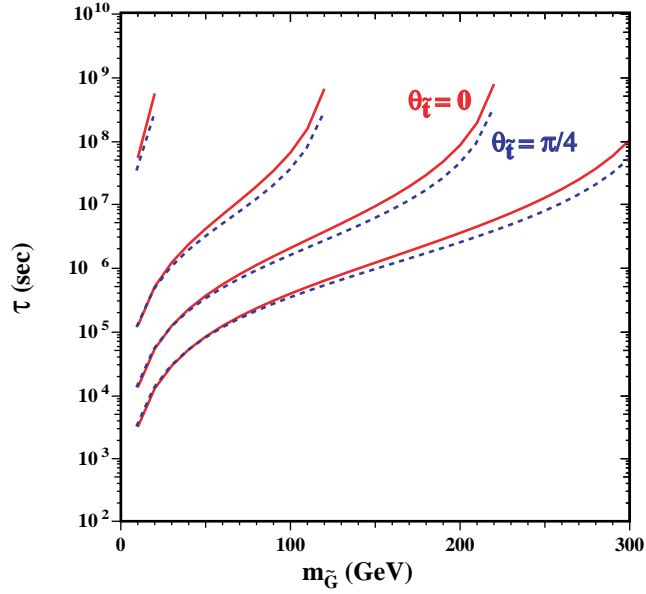,height=8cm}}
\end{center}
\caption{\label{fig:2body}\it
The stop lifetime as a function of $m_{\widetilde G}$ for $\mstop = 200, 300, 400$ and
$500$~GeV (top to bottom), shown for the case of the two-body decay ${\tilde t_1} \to {\widetilde G} t$,
the dominant mode for $m_{\widetilde G} < \mstop - m_t$, assuming zero stop mixing (red solid line) 
and maximal mixing (blue dashed line).}
\end{figure}

\item Case 2: $m_W + m_b < \Delta m < m_t$. In this case, the dominant decays are
into the three-body final state $\stop_1 \to \tilde{G} + W + b$. We identify three tree-level
decay diagrams, proceeding via $t$, ${\tilde b}$ and chargino exchange. The amplitudes are
\bea
{\cal{M}}_t &= &C_t P_t(q_1)  {\bar{\Psi}}_\mu  \, p^{\mu} [ A_{\ts} + B_{\ts} \gamma_5 ] 
                          (\slash{q}_1 + m_t) \gamma^{\rho} \epsilon_{\rho}(k)
			  P_L \, v(p_2) , \\
{\cal M}_{\sbot_i} &=& C_{\bs_i} P_{\bs_i}(q_2)  {\bar{\Psi}}_\mu  \, q^{\mu}_2 [ a_i P_L + b_i P_R  ] 
                                           \,   p^{\rho} \epsilon_{\rho}(k) \, v(p_2) , \\
{\cal M}_{\chi^+_i} &=& C_{\chi^+} P_{\chi^+_i}(q_3)  {\bar{\Psi}}_\mu \gamma^{\rho}  
\epsilon_{\rho}(k)\gamma^\mu 
                                     [ V_i + A_i \gamma_5 ] 
            (\slash{q}_3 + m_{\chi} )  [ S_i + P_i \gamma_5 ]  v(p_2) ,
\eea 
where $C_t ={g_2}/{2 M}$, $C_{\bs_i}={2g_2 \kappa_i}/{ M}$, $C_{\chi^+}={m_W}/{ M}$, 
and $p$ is the initial stop four-momentum. We define
$q_1 \equiv p - p_1$, $q_2 \equiv p - k$ and $q_3 \equiv p - p_2$, with $p_1, k, p_2$ denoting the outgoing
four-momenta of the gravitino, $W$ boson, and $b$ quark respectively,
and $\epsilon_{\rho}(k)$ denotes  the W polarization vector. 
Expressions for  $A_{\ts}, B_{\ts}, a_i, b_i,  \kappa_i, V_i, A_i, S_i$ and $P_i$ are presented in the Appendix.

Squaring and summing over final polarizations we obtain:
\beq
|{\bar{\cal{M}}}|^2= |{\cal{M}}_t|^2+  |{\cal{M}}_{\bs}|^2+|{\cal{M}}_{\chi^+}|^2
     + 2 \, {\rm Re}[ {\cal{M}}^\dagger_t {\cal{M}}_{\bs}+ {\cal{M}}^\dagger_t
     {\cal{M}}_{\chi^+}+ 
            {\cal{M}}^\dagger_{\bs} {\cal{M}}_{\chi^+}].
\eeq
where the sums over sbottom and chargino indices are implicit. 
The individual squared amplitudes can be written as:
\beq
|{\cal{M}}_{\psi_a}|^2 = C^2_{\psi_a} \, |P_{\psi_a}(q_a)|^2  \, W_{\psi_a\psi_a } ,
\eeq
where $\psi_a=(t,\bs_j, \chi^+_k)$,  and the functions $W_{\psi_a\psi_a}$ are functions of
the scalar products of the momenta $p,p_1,p_2,k$.
The interference terms may be written as follows:
\beq
 {\cal{M}}^\dagger_{\psi_a} {\cal{M}}_{\psi_b} = C^*_{\psi_a}  C_{\psi_b}   
                       P^*_{\psi_a} (q_a) P_{\psi_b} (q_b) 
                                                  W_{\psi_a\psi_b },
\eeq
where the functions $ W_{\psi_a\psi_b}$ can be written also in terms of the invariants.
The functions $P_{\psi_a}  (q_a)$ are propagator factors, e.g., for the top
quark ${\psi_a}= t$, and we have
\beq
P_t(q_1)=\frac{1}{q^2_1-m^2_t+i \epsilon}.
\eeq
There are similar expressions for the sbottom and chargino  contributions,
$P_{\bs}(q_2)$ and $P_{\chi^+}(q_3)$ respectively. 

Detailed formulae for the
functions $W_{\psi_a\psi_a}$ and $ W_{\psi_a\psi_b}$ are given in the Appendix.
Using these, we calculate the decay width:
\beq
\frac{d\Gamma}{dx\, dy} = \frac{ m^2_{{\ts}_1} }{256 \, \pi^3 } |{\bar{\cal{M}}}|^2 ,
\eeq
where the integration limits are (in the limit when we neglect the
bottom quark mass): $2\mu_G < x < 1+\mu_G-\mu_W$ and $y_{-} < y < y_{+}$,
where $\mu_i = m^2_i / m^2_{{\ts}_1}$ and:
\beq
y_{\pm} = \frac{ 1+\mu_G+\mu_W -x}{2( 1+\mu_G -x)} 
           [ (2-x)\pm (x-4\mu_G)^{1/2} ]  .
\eeq  
Integrating this equation numerically, we find that in
significant regions of parameter space the light stop
has a lifetime of order $10^{9}-10^{14}$~s or more.
Typical results are shown in
Fig.~\ref{fig:3body}. As expected, the typical lifetimes are orders
of magnitude longer than those in Fig.~\ref{fig:2body}. 
They are also sensitive
mainly to the sparticle masses, and relatively insensitive to the stop mixing angle, as
can be seen by comparing the red solid and blue dashed curves, as well as being insensitive
to the sbottom mixing angle $\theta_{\tilde b}$, which is assumed here to 
vanish~\footnote{Typical values
in the allowed NUHM models discussed later are $\theta_{\tilde b} \sim 0.04$.}.

\begin{figure}
\begin{center}
\mbox{\epsfig{file=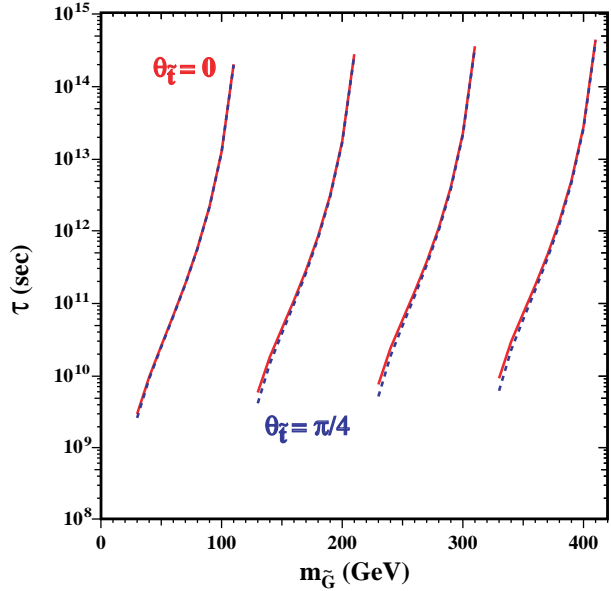,height=8cm}}
\end{center}
\caption{\label{fig:3body}\it
The stop lifetime as a function of $m_{\widetilde G}$ for $\mstop = 200, 300, 400$ and
$500$~GeV, shown for the case of three-body decays ${\tilde t_1} \to {\widetilde G} W b$,
the dominant mode for $m_W + m_b < \Delta m < m_t$, assuming zero stop mixing (red solid line)
or maximal mixing (blue dashed line).}
\end{figure}

\item Case 3:  $m_b + \Lambda_{QCD} < \Delta m < m_W + m_b$. 
In this case, the real $W$ of the previous case must become virtual, and the dominant decays
are four-body: ${\tilde t_1} \to {\widetilde G} + b + {\bar q}q$ or $\ell \nu$. The decay rate for this 
case is further suppressed compared to Case 2, and we estimate that
the stop lifetime would in this case exceed $10^{12}$~s. Thus, the stop might even decay
after the release of the CMB, in which case there would be important constraints from the
absence of distortions in CMB spectrum. We have not explored this issue, in view of the 
relatively small region of parameter space concerned.

\end{enumerate}

It is apparent from the above discussion that not only does the stop live long enough to
hadronize and pass through collider detectors, but it may also live long enough to wreak
cosmological havoc. We discuss each of these aspects in the following Sections.

\section{Spectroscopy of Stop Hadrons and their Decays}

The metastable stop would hadronize to produce both $\stop_1 qq$ `sbaryons'
and $\stop_1 {\bar q}$ `mesinos' and their antiparticles, many of whose aspects are
discussed in~\cite{GL}. On general QCD
principles and by analogy with the spectroscopy of charmed hadrons, one expects the
${\widetilde T}^0 \equiv \stop_1 {\bar u}$, ${\widetilde T}^+ \equiv \stop_1 {\bar d}$ and 
${\widetilde T}_s \equiv \stop_1 {\bar s}$ mesinos to be the lightest
stop hadrons. As was pointed out in~\cite{GL}, one can expect the ${\widetilde T}^0$
mesino and its antiparticle to be strongly mixed.
Since $m_s > m_d > m_u$, and since the ${\widetilde T}^+$ 
and ${\widetilde T}_s$ mesinos, being charged, would acquire 
additional electromagnetic
mass corrections, we expect $m_{{\widetilde T}_s} - m_{{\widetilde T}^+} $
and $m_{{\widetilde T}^+} - m_{{\widetilde T}^0}$ to be similar to the measured values of the
$D_s - D^+$ and $D^+ - D^0$ mass differences, namely $\simeq 99$~MeV and
$\simeq 4.8$~MeV, respectively~\cite{PDG}. 

Correspondingly, we would expect the ${\widetilde T}_s$ mesino to decay weakly
into ${\widetilde T}^0 e \nu$ with a lifetime similar to that of the muon,
namely $\simeq 2 \times 10^{-6}$~s, and the ${\widetilde T}^+$ 
mesino to decay weakly into ${\widetilde T}^0 e \nu$ with a lifetime
\begin{equation}
\tau_{{\widetilde T}^+} \simeq \tau_n \times \left(
\frac{m_n - m_p}{m_{{\widetilde T}^+} - m_{{\widetilde T}^0}} \right)^5 \simeq 1.2~{\rm s} .
\label{mesinolife}
\end{equation}
These mesino lifetimes are also such that they would pass through a typical collider
detector before decaying. In the early Universe, the ${\widetilde T}_s$ would have
decayed very quickly after being formed at the quark-hadron transition, whereas the
${\widetilde T}^+$, if they survive, would have decayed near the beginning of BBN, 
and so would not have
affected its end results. Moreover, since the stable ${\widetilde T}^0$
mesino would be neutral, it could not have catalyzed light-element nucleosynthesis by
bound-state effects. The only potential cosmological danger from the mesinos
would be the supersymmetric decays of the ${\widetilde T}^0$ into the ${\widetilde G}$
and conventional particles, as discussed in~\cite{CEFOS}, for example.

Turning now to the $\stop_1$ sbaryons, again by analogy with the charmed baryons,
we would expect the lightest state to be the $\Lambda_{\widetilde T}^+
\equiv {\tilde t_1} ud$, with the other
sbaryons $\Sigma_{\widetilde T}^{++,+,0} \equiv \stop_1 (uu, ud, dd)$, 
$\Xi_{\widetilde T}^{+,0} \equiv \stop_1 s (u,d)$ being heavier by amounts $\sim
\Lambda_{QCD}, m_s - m_{d,u}$, respectively. Just like the ${\widetilde T}_s$ mesino
discussed above, these heavier sbaryons would have decayed innocuously before BBN. For
example, if the $\Sigma_{\widetilde T} - \Lambda_{\widetilde T}$ mass difference were
similar to the corresponding mass differences among charmed and bottom baryons, 
namely $\sim 170$~\cite{PDG} to 190~MeV~\cite{sigmab}, the $\Sigma_{\widetilde T}$
would decay very rapidly via the strong interactions. If the $\Xi_{\widetilde T} - \Lambda_{\widetilde T}$ mass difference were similar to the corresponding mass difference among charmed baryons, namely $\sim 180$~MeV, the $\Xi_{\widetilde T}^0 = \stop_1 sd$ state would decay
semileptonically with a lifetime $< 10^{-6}$~s into the $\Lambda_{\widetilde T}$, whereas the
$\Xi_{\widetilde T}^+ = \stop_1 su$ state would decay semileptonically into the 
$\Sigma_{\widetilde T}^{++} = \stop_1 uu$ with a longer lifetime $\sim 10^{-2}$~s (because of the
much smaller phase space $\sim 15$~MeV for the decay). However, this decay would also occur by the
beginning of BBN.
Thus, the cosmological dangers could potentially arise only from the supersymmetric decays of the
$\Lambda_{\widetilde T}^+$ and its antiparticle, the ${\bar \Lambda}_{\widetilde T}^-$,
and dangerous catalysis effects could only be due to bound states of the 
${\bar \Lambda}_{\widetilde T}^-$.

We note in passing some similarities with and differences from the case of a metastable charge -1/3
squark, such as the lighter sbottom, ${\tilde b_1}$. In this case, we would expect the lightest
sbaryon to be the $\Lambda_{\widetilde B}^0 \equiv {\tilde b_1} ud$. Its decays might cause
cosmological problems, but it could not cause dangerous bound-state effects. The nature of
the lightest sbottom mesino is not so clear. The fact that the $d$ quark is heavier than the
$u$ quark would tend to make the ${\widetilde B}^0$ mesino heavier, but the electromagnetic
corrections would add to the mass of the ${\widetilde B}^+$. Experimentally, the situation
with $B$ mesons is ambiguous, $m_{B^0} - m_{B^+} = 0.33 \pm 0.28$~MeV. However, the
likelihood is probably that the ${\widetilde B}^+$ would be lighter, in which case its antiparticle, the
${\widetilde B}^-$, would generate bound-state effects, unlike the ${\widetilde T}^0$.

\section{Collider Lower Limit on the Stop Mass}

As we have seen, the stop would have a very long lifetime in GDM scenarios of
the type considered here, and would hadronize before passing through a typical
collider detector. The relative production rates of mesons and baryons containing
heavy quarks are not established, and neither are the relative production rates of
heavy-quark mesons containing strange quarks. We assume for simplicity that
half of the produced stops hadronize into charged mesinos or sbaryons, and half
into neutral stop hadrons. These would be produced embedded within hadronic jets, but 
conventional QCD fragmentation ideas suggest that the stop hadrons would carry
essentially all the energies in these jets, with energy fractions 
$z_{\widetilde T} \sim 1 - \Lambda_{QCD}/m_{\tilde t_1}$. 

The typical energy loss
as the stop hadron passes through a detector tracking system would be very small.
There would also be nuclear interactions, particularly in calorimeters. In addition to the 
familiar charge-exchange reactions, these would also include baryon-exchange
reactions, whereby a stop mesino striking a nucleus would convert into a stop
sbaryon: ${\widetilde T} + (p,n) \to (\Lambda_{\widetilde T}, \Sigma_{\widetilde T}) + n \pi$,
whereas the corresponding sbaryon-to-mesino conversion would be essentially forbidden.
It has been pointed out that the baryon-exchange process would be almost 100~\% efficient 
in converting heavy
mesinos to sbaryons when they traverse material with a thickness of 1~m or more~\cite{GL}
(see also~\cite{ADPRW}).
We therefore assume for simplicity that all of the stop hadrons emerging from calorimeters 
into muon detectors are
sbaryons, that half of them are singly-charged, and that this charge is  independent of the
charge of the stop hadron at production. This would imply that just a quarter of the produced
stop hadrons would be singly-charged both at production and in the muon detectors. Thus,
only about 1/16 of the produced stop-antistop pairs would yield a robust
signature of a pair of oppositely-charged massive metastable particles.
 
We use here the limits set by
direct searches for the pair-production of massive (meta)stable charged particles at hadron colliders to
set a lower bound on the stop mass. Nunnemann~\cite{nunnemann} gives an upper
limit from D0 of about 0.1~pb from a search for the pair-production of massive 
oppositely-charged particles, and a similar upper limit for the pair-production of
stops has been presented by CDF~\cite{Tevatron}, which is used to set
a lower limit of 220~GeV for
$\mstop$. Gallo~\cite{gallo} gives a D0 upper limit on the production of
neutral gluino hadrons of about 0.5~pb. Again assuming that about a quarter of the
stop hadrons are produced neutral and also appear neutral in the outer detectors, 
this limit gives a somewhat
weaker limit on $\mstop$. Therefore, we assume $\mstop > 220$~GeV~\cite{Tevatron}.

\section{Stop NLSP in the CMSSM}

We now discuss the prospects for finding a stop NLSP in the CMSSM, i.e., the simplest 
variant of the MSSM, in which the soft supersymmetry-breaking masses are
universal at the GUT scale. Thus, we have as free parameters $m_0$, the universal soft scalar mass at
the GUT scale, $m_{1/2}$, the universal gaugino mass at the GUT scale, $A_0$,
the universal trilinear soft supersymmetry-breaking parameter at the GUT scale, 
$\tan \beta$, the ratio of the two MSSM Higgs vevs,
and the sign of $\mu$ (where $\mu$ is the Higgs mixing parameter). 
In addition, unless we make additional assumptions, as may be motivated by
supergravity models, we must consider the gravitino mass $m_{\widetilde G}$ as an extra
free parameter, which is chosen so that the gravitino is the lightest
supersymmetric particle (LSP). 

We search within the CMSSM for a set of
parameters where not only is the stop the NLSP, but also all the known experimental
and phenomenological constraints on supersymmetry are satisfied, including the $b \to s \gamma$
decay rate~\cite{bsg,bsgth}, and the LEP lower bounds on the masses of the chargino
and the Higgs boson~\cite{mh}~\footnote{We use the public Fortran code
{\tt FeynHiggs}~\cite{FH} to calculate $m_h$.}. In view of the ambiguity in the value of the hadronic 
contribution to the Standard Model value of the muon anomalous magnetic moment,
$g_\mu-2$, we omit this observable  for the
moment. The $B_s \to \mu^+ \mu^-$ constraint is significant only for large
$\tan \beta$, whereas, as we see below, the regions that are relevant to our search
have relatively small $\tan \beta \sim 10$.

We choose the sign of $\mu$ to be positive, as our search indicates that negative
$\mu$ has less chance of yielding a stop NLSP~\footnote{Negative $\mu$ may also be
disfavoured by $b \to s \gamma$ and $g_\mu - 2$.}, and assume $m_t =
171.4$~GeV~\cite{mt171p4}  and $m_b(m_b)^{\overline{MS}} = 4.25$~GeV. 
Since $m_\chi \simeq 0.43 m_{1/2}$, in order to obtain $m_\chi > m_{\tilde t_1} > 220$ GeV,
we must set $m_{1/2} \ga 520$~GeV in the CMSSM (this is relatively independent of $\tan \beta$).
However, the LL and RR components of the $m_{\ts}^2$ mass matrix
receive contributions of about $6 m_{1/2}^2$ and $4 m_{1/2}^2$ respectively,
forcing one to consider
large off-diagonal elements. These are of the form $m_t X_t = -m_t (A_t +
\mu/\tan \beta)$, as seen in (2)~\footnote{Note that $A_t$
signifies the value of the trilinear term at the scale $m_{\ts}$, which differs from its value $A_0$
at the GUT input scale.}.
Low stop mass eigenvalues therefore require a combination of 
high values of $A_0$ and relatively low values of $\tan \beta$.  
In fact, as we show explicitly below, only intermediate values of $\tan \beta$ have
any chance of realizing low stop masses without upsetting 
the remaining phenomenological constraints.  At high $ \tan \beta$
the $b \to s \gamma$ constraint is not satisfied, and at low $\tan \beta$ the 
LEP Higgs mass bound
cannot be satisfied.  Our only option therefore is large $A_0$.
Our search in the CMSSM is further complicated by the dependence of the Higgs mass on $X_t$.
For relatively small values of $|X_t|/m_{1/2}$, the Higgs mass increases with increasing 
$|X_t|$.  However, for $|X_t|/m_{1/2} \ga 2$, the Higgs mass begins to decrease rapidly~\cite{xhiggs}.
In order to obtain a light stop, we need $|X_t|/m_{1/2} \sim 6 m_{1/2}/m_t$,
which is too large to have any chance of satisfying the Higgs mass constraint.

\begin{figure}
\begin{center}
\vspace{-1in}
\mbox{\epsfig{file=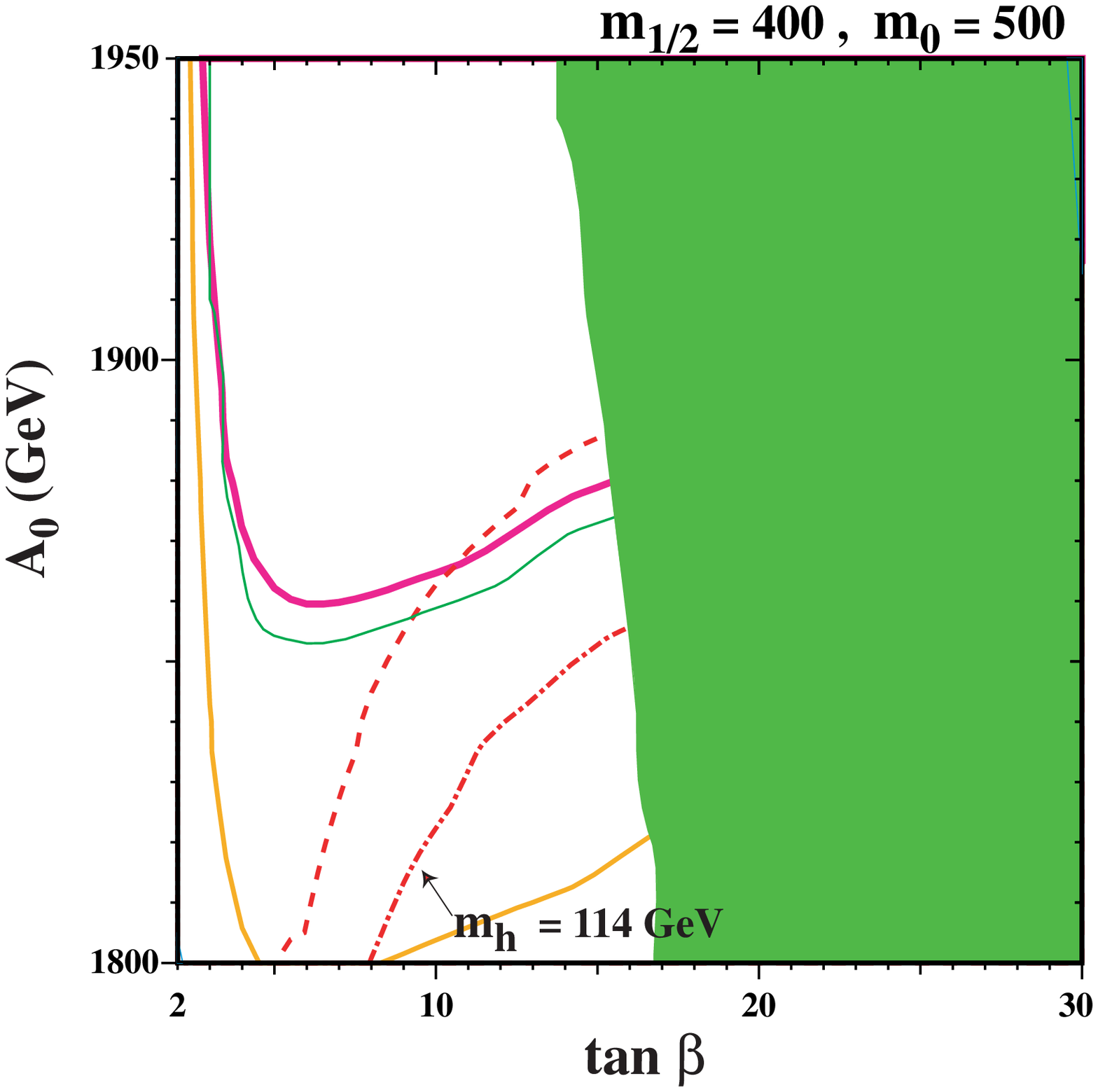,height=8cm}}
\mbox{\epsfig{file=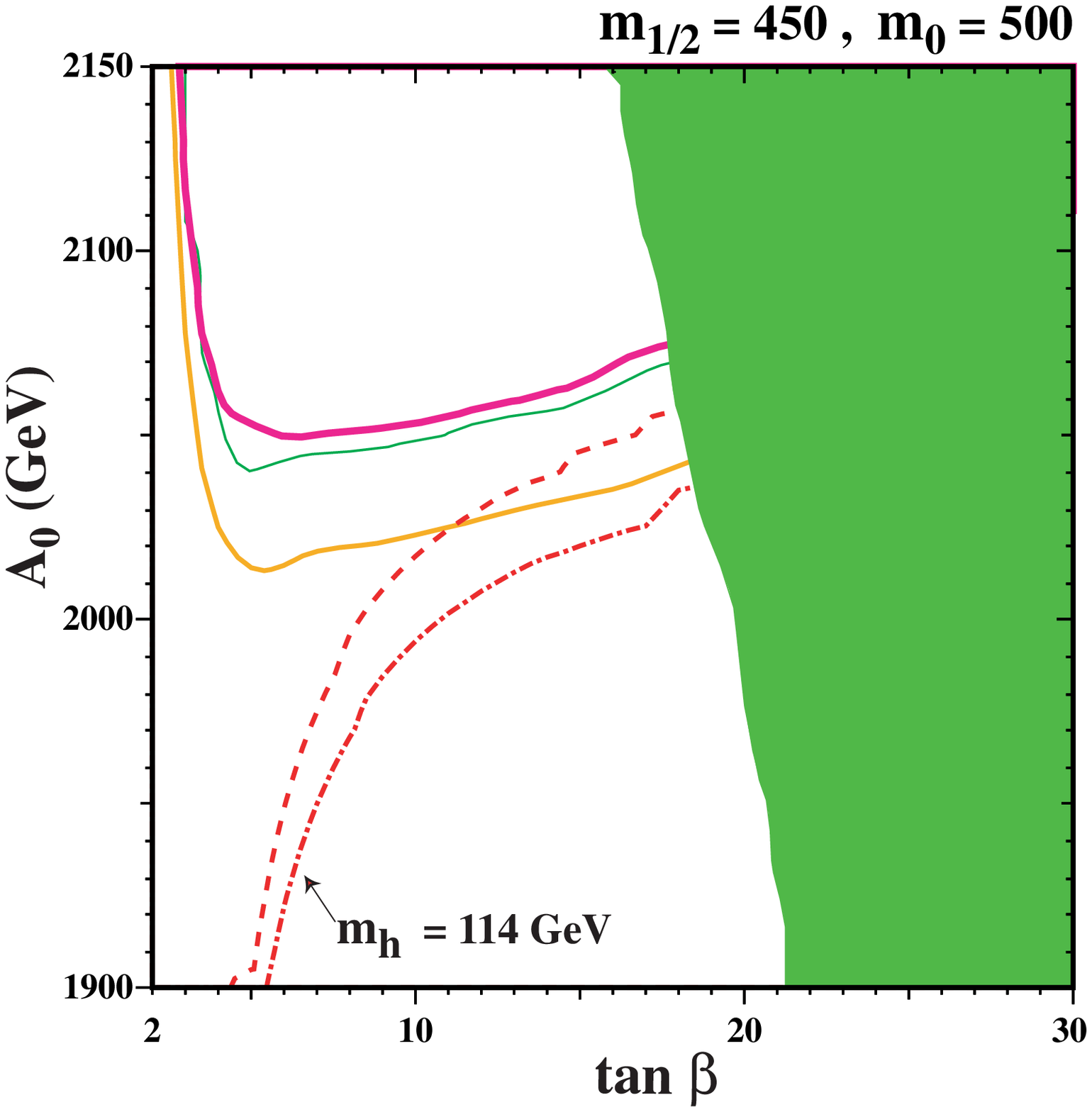,height=8cm}}
\end{center}
\begin{center}
\mbox{\epsfig{file=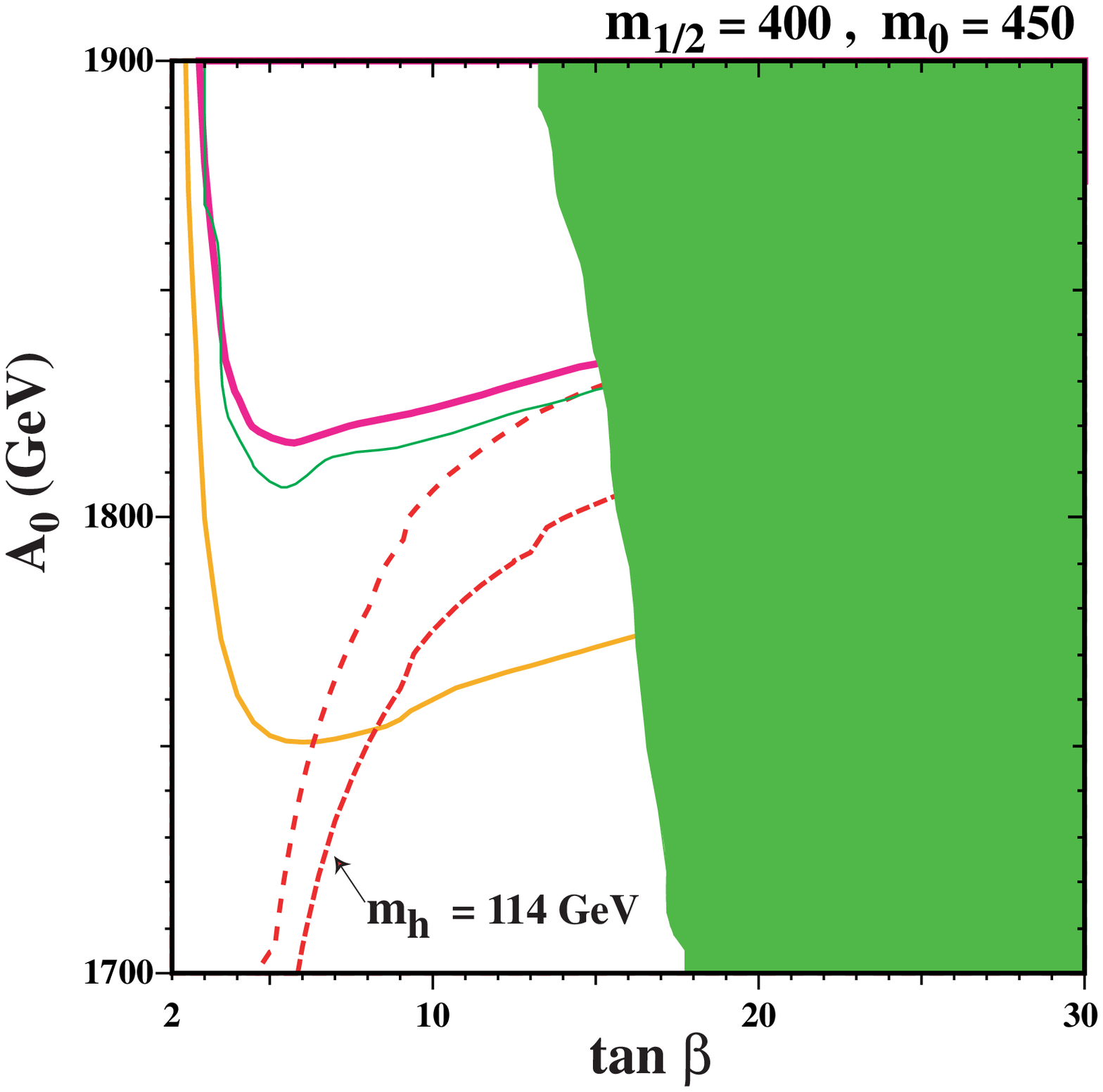,height=8cm}}
\mbox{\epsfig{file=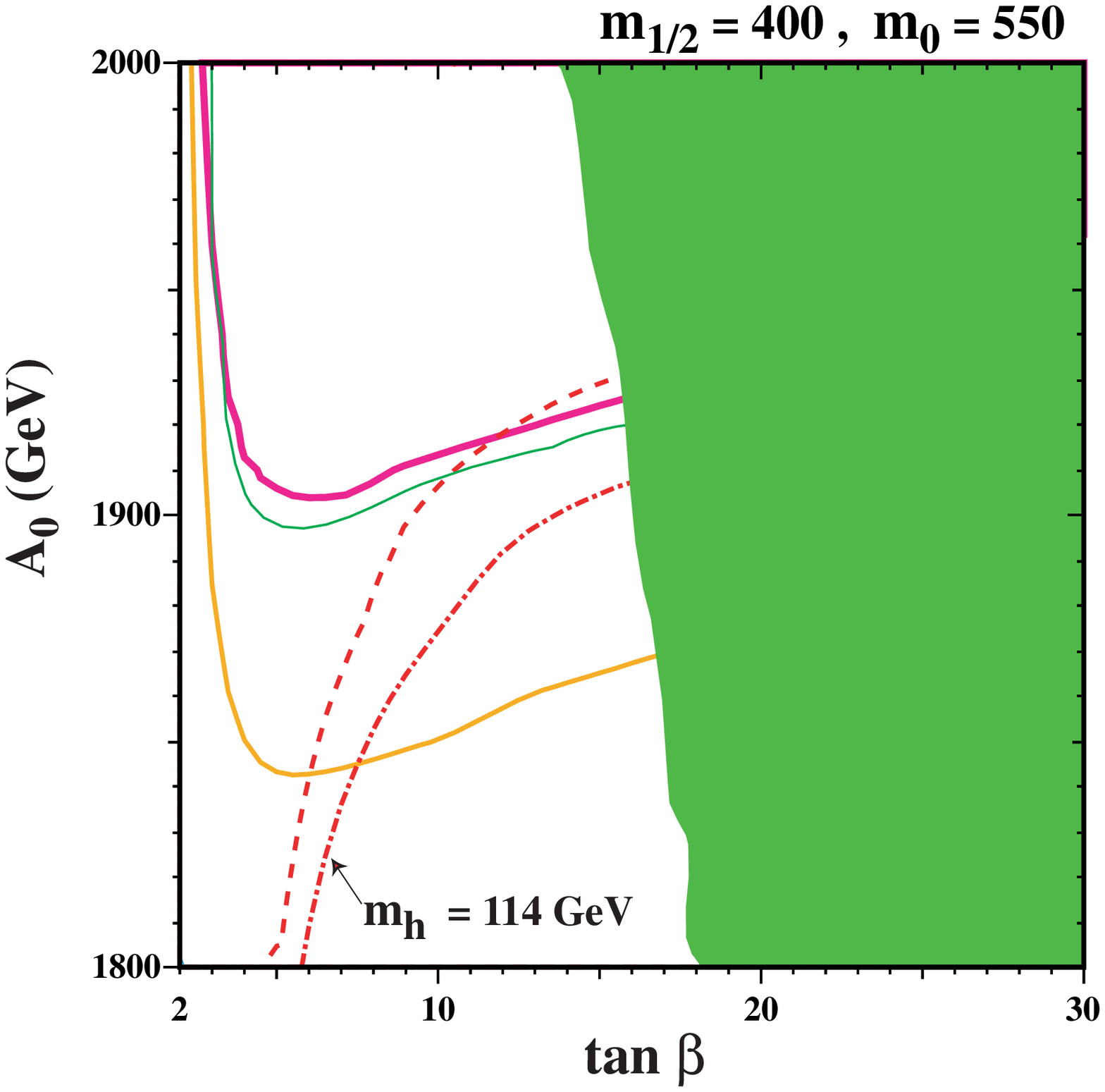,height=8cm}}
\end{center}
\caption{\label{fig:cmssm1}\it
The $(\tan \beta, A_0)$ plane in the CMSSM, for $(m_{1/2},m_0)=$ (a)
(400,500)~GeV, (b) (450,500)~GeV, (c) (400,450)~GeV, and (d) (400,550)~GeV
respectively. We use $m_t = 171.4$~GeV,
$m_b(m_b)^{\overline{MS}} = 4.25$~GeV, and $\mu > 0$. 
The neutralino stop degeneracy contour is plotted as the thick solid purple line:
above this line stop is lighter and the NLSP, assuming a light gravitino. 
The solid orange line is the contour for $m_{\tilde{t}_1} = 220$~GeV: above this line,
the stop is too light. 
Large $\tan \beta$ is excluded by the $b \to s \gamma$ constraint (green shaded
region). The Higgs likelihood exclusion line (preferred) is drawn as a dashed red line, while
the face value $m_h = 114.4$~GeV (deprecated) is the dot-dashed line.  
Also plotted as the thin green line is the stop relic density 
$\Omega_{\stop} h^2 = 4\times 10^{-4}$. 
}
\end{figure}

In order to obtain a more
comprehensive picture, we show in Fig.~\ref{fig:cmssm1} some contour plots in 
$(\tan \beta, A_0)$ planes for some fixed values of $m_0$ and $m_{1/2}$.
In panel (a) we fix $m_0 = 500$~GeV and $m_{1/2} =
400$~GeV (the gravitino mass $m_{\widetilde G}$ is irrelevant here). 
The $b \to s \gamma$ constraint, which excludes large $\tan \beta$, is shown by the green shaded
region.
In the allowed region, either the stop or the neutralino is the lightest sparticle in the
spectrum. Above
the thick purple solid line, the stop is lighter, while below it the neutralino is lighter. We
also plot the $\mstop = 220$~GeV contour, represented by the orange solid line:
regions above this line have $\mstop < 220$~GeV, and therefore are excluded.  
The Higgs mass constraint is represented by two red lines, the dashed line is based on
a likelihood analysis, and the dot-dashed line on the face value of the Higgs mass limit,
namely $m_h = 114.4$~GeV (deprecated). The constraints are satisfied only below the lines. 
We use the constraint determined by the likelihood function in our analysis.
We see in panel (a) that there is an overlap between the stop NLSP
region and the region allowed by the Higgs likelihood constraint. 
However, the stop mass is around 150-160~GeV in this region, which is therefore 
excluded by the lower bound on the stop mass. We also plot in
Fig.~\ref{fig:cmssm1}(a), the contour where the 
NLSP would have a relic density of $\Omega_{\tilde t} h^2 = 4 \times 10^{-4}$, if it did not decay.
This is shown by the thin green line that lies below the neutralino-stop contour.
The small value is a consequence of the strong stop-antistop annihilation cross section.

When we increase $m_{1/2}$, for example to 450~GeV as in panel (b), both
$m_\chi$ and $\mstop$ are raised and neutralino-stop degeneracy is reached
at higher $A_0$. We see that the $\mstop = 220$~GeV line is
closer to the neutralino-stop degeneracy line, and if we kept increasing
$m_{1/2}$ we would be able to find points where the stop is lighter than the neutralino and
has mass larger than 220~GeV. However, although the Higgs mass constraint also
moves to higher $A_0$, it moves slower than the previous two lines. As a result,
there is no overlap region where there is a stop NLSP and the Higgs mass
bound is satisfied.

The first two panels in Fig~\ref{fig:cmssm1} already suggest that there are no
allowed regions with a stop NLSP and gravitino LSP in the CMSSM.  
Generalizing this  observation, we first note that, due to the nature of the 
RGEs in the MSSM, varying $m_0$ would have less effect than varying $m_{1/2}$. This is shown
explicitly by panels (c) and (d) of Fig.~\ref{fig:cmssm1},  which have the same
$m_{1/2}=400$~GeV as in panel (a) but with $m_0 = 450, 550$~GeV respectively.
Going from panel (a) to panel (c), the decrease in $m_0$ results in lower $m_h$
and there is no longer any overlap between the region allowed by $m_h$ and
the region where the stop is the NLSP
With higher $m_0$ as in panel (d), we get heavier $\stop_L$ and
$\stop_R$, and hence we need higher $X_t$ and hence $A_0$
in order to approach $\stop_1 - \chi$ degeneracy.
However, higher $X_t$ in turn has a problem with the $m_h$ constraint as
described above.  This illustrates our conclusion that, indeed,
a stop NLSP scenario is not possible within the CMSSM. 

\section{Stop NLSP in the NUHM}

We next study the MSSM with non-universal Higgs masses (NUHM)~\cite{ourNUHM}. 
In the NUHM, the soft supersymmetry-breaking masses in the Higgs sector, 
$m_1$ and $m_2$, are not
necessarily equal to the sfermion soft mass $m_0$ at the GUT scale. Using the
radiative electroweak symmetry breaking  conditions, we can characterize the
new parameters as $\mu$ and $m_A$ (the pseudoscalar Higgs mass), both values
being defined at the weak scale:
\begin{equation}
\mu^2 \; = \; \frac{m_1^2 + m_2^2 \tan^2 \beta + \frac{1}{2}(1 - \tan^2 
\beta) 
+ \Delta_\mu^{(1)}}{\tan^2 \beta - 1 + \Delta_\mu^{(2)}} ,
\label{vac1}
\end{equation}
and
\begin{equation}
m_A^2(Q) \; = \; m_1^2(Q) + m_2^2(Q) + 2 \mu^2(Q) + \Delta_A(Q).
\label{vac2}
\end{equation}
Many different sparticles could be the NLSP in this model, in different regions
of the NUHM parameter space. These include the lightest neutralino $\chi$, the lighter
stau $\tilde{\tau}_1$, the selectron (smuon) $\tilde{e}_R$ ($\tilde{\mu}_R$), the lighter
stop $\tilde{t}_1$, the up squark (charm squark) $\tilde{u}_R$ ($\tilde{c}_R$),
and the tau sneutrino $\tilde{\nu}_{\tau}$. 
Thanks to the RGE, the sbottom tends to be heavier than the stau, so unless we
have
non-universal
soft scalar supersymmetry-breaking masses for squarks and sleptons at the GUT
scale, sbottom could not be the NLSP.  
The up and charm squarks
could be the NLSP  for very large $|\mu|$ and $m_A$~\cite{baerNUHM}, where the Higgs
soft mass-squared $m_1^2$ or $m_2^2$ become negative at the GUT scale. 
However, we discard
this possibility, preferring to impose a GUT stability
constraint~\cite{ourNUHM}. We also do not
consider the other possible NLSPs in this paper, and focus only on the stop.  

\begin{figure}
\begin{center}
\mbox{\epsfig{file=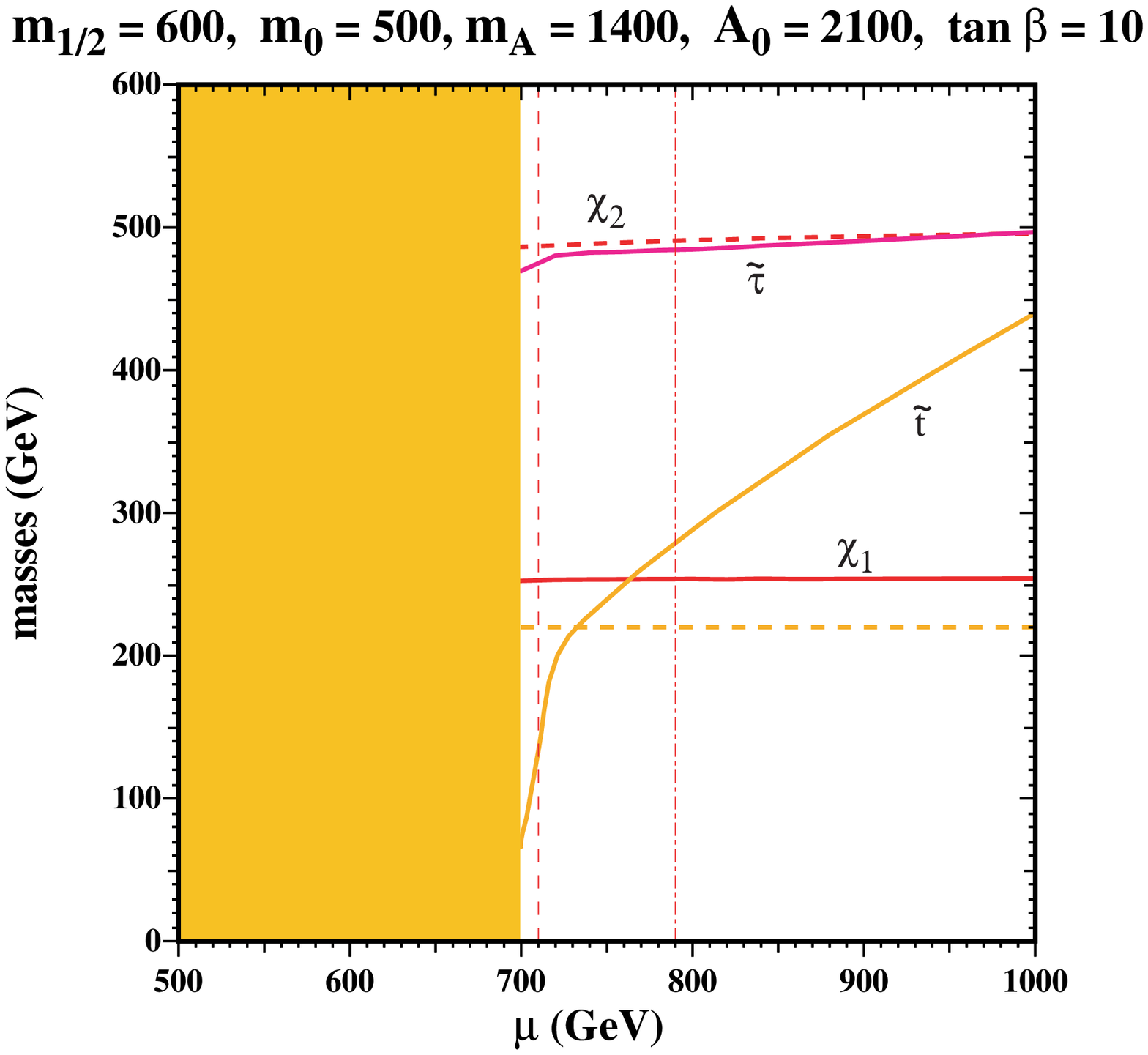,
height=7cm}}
\mbox{\epsfig{file=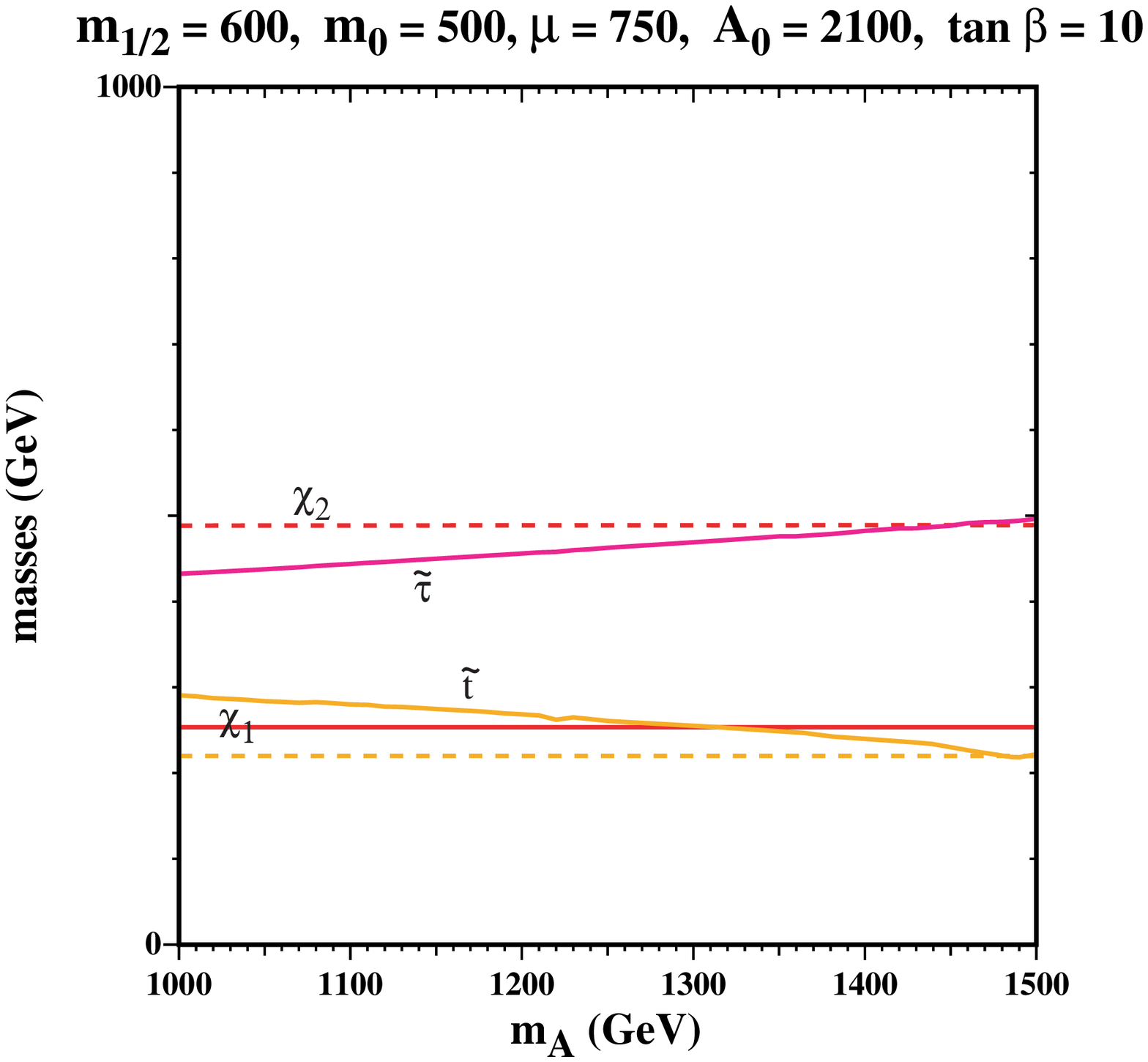,
height=7cm}}
\end{center}
\caption{\label{fig:nuhm1}\it
Sparticle masses in the NUHM as functions of (a) $\mu$ with $m_A = 1400$~GeV,
and (b) $m_A$ with $\mu = 750$~GeV, both with $\tan \beta = 10$,
$A_0 = 2100$~GeV, $m_0 = 500$~GeV and $m_{1/2} = 600$~GeV. We plot the masses of
the lightest neutralino (solid red), the second lightest neutralino (dashed red),
the lighter stau (purple solid), and the lighter stop (orange solid). In the shaded region,
the stop becomes tachyonic. The
Higgs likelihood constraint is shown by the vertical dashed red line. We also draw a
horizontal dashed line at 220~GeV to indicate the lower bound on the stop mass.}
\end{figure}

In order to obtain a light stop, small $|\mu|$ is preferred and, more weakly, large $m_A$. In
Fig.~\ref{fig:nuhm1}(a) we plot various masses as functions of $\mu$ for the fixed values
$m_A = 1400$~GeV, $\tan \beta = 10$,
$A_0 = 2100$~GeV, $m_0 = 500$~GeV (in the region of the values studied previously
within the CMSSM) and $m_{1/2} = 600$~GeV (close to the minimum allowed so that
$m_\chi > m_{\stop_1} > 220$~GeV). We see that there is
a small range $\mu \simeq 730 - 770$~GeV where the stop is the NLSP and survives the
phenomenological constraints. The stop relic density in this region is 
$\Omega_{\stop_1} h^2 \sim
10^{-4}$, and the neutralino mass is about 250~GeV. We note that this region is nearly excluded by the Higgs mass constraint.
Indeed, if we had taken the LEP limit on $m_h$ at its face value, this region would have been
excluded.  Panel (b) shows the spectrum
as a function of $m_A$, for $\mu = 750$~GeV. Here we see that $\mstop$ is a decreasing
function of $m_A$, and there is a region where a stop NLSP is allowed, at $m_A \simeq 1350  -
1500$~GeV.  

\begin{figure}
\begin{center}
\vspace{-1in}
\mbox{\epsfig{file=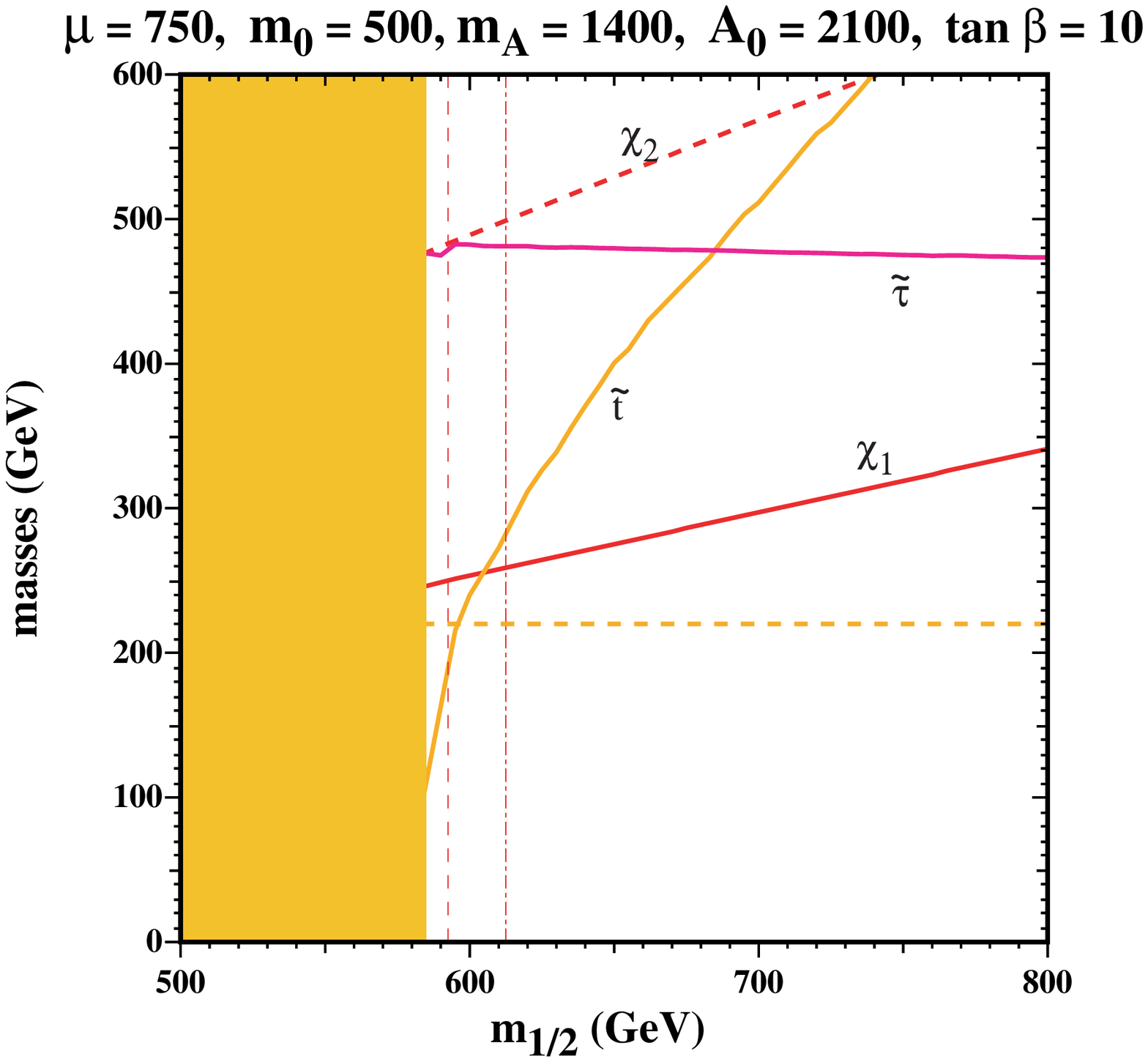,height=7cm}}
\mbox{\epsfig{file=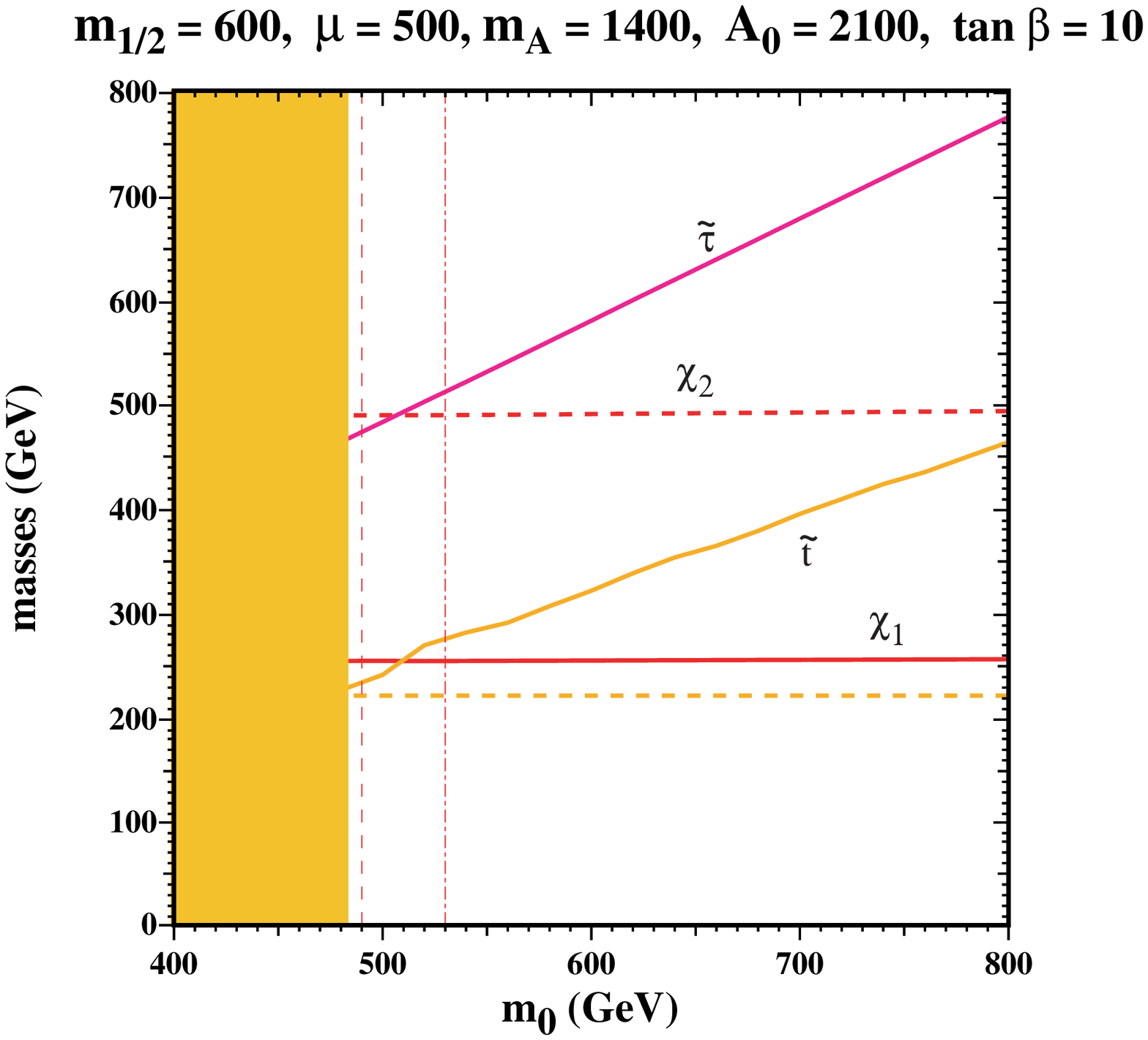,height=7cm}}
\end{center}
\begin{center}
\mbox{\epsfig{file=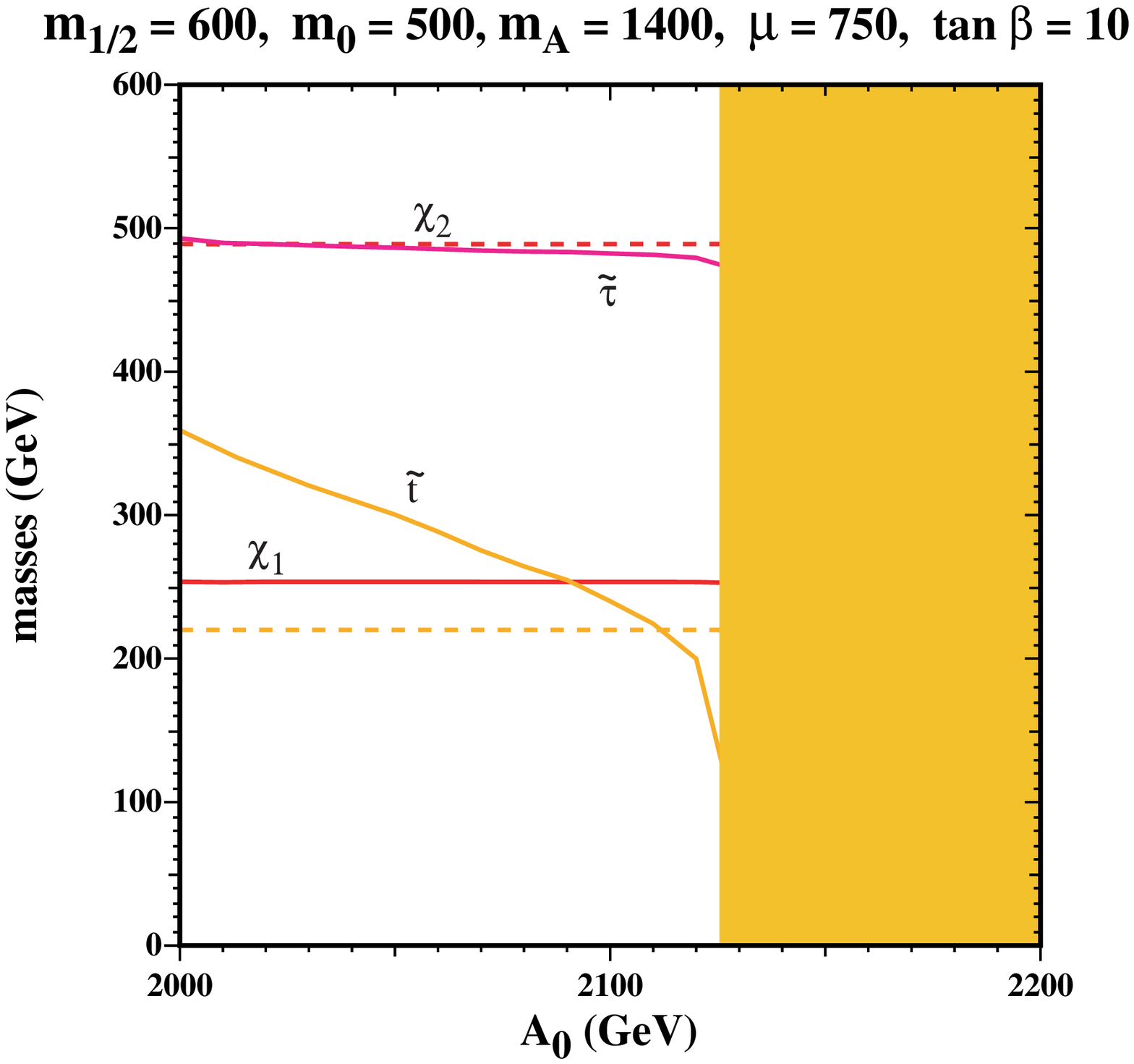,height=7cm}}
\mbox{\epsfig{file=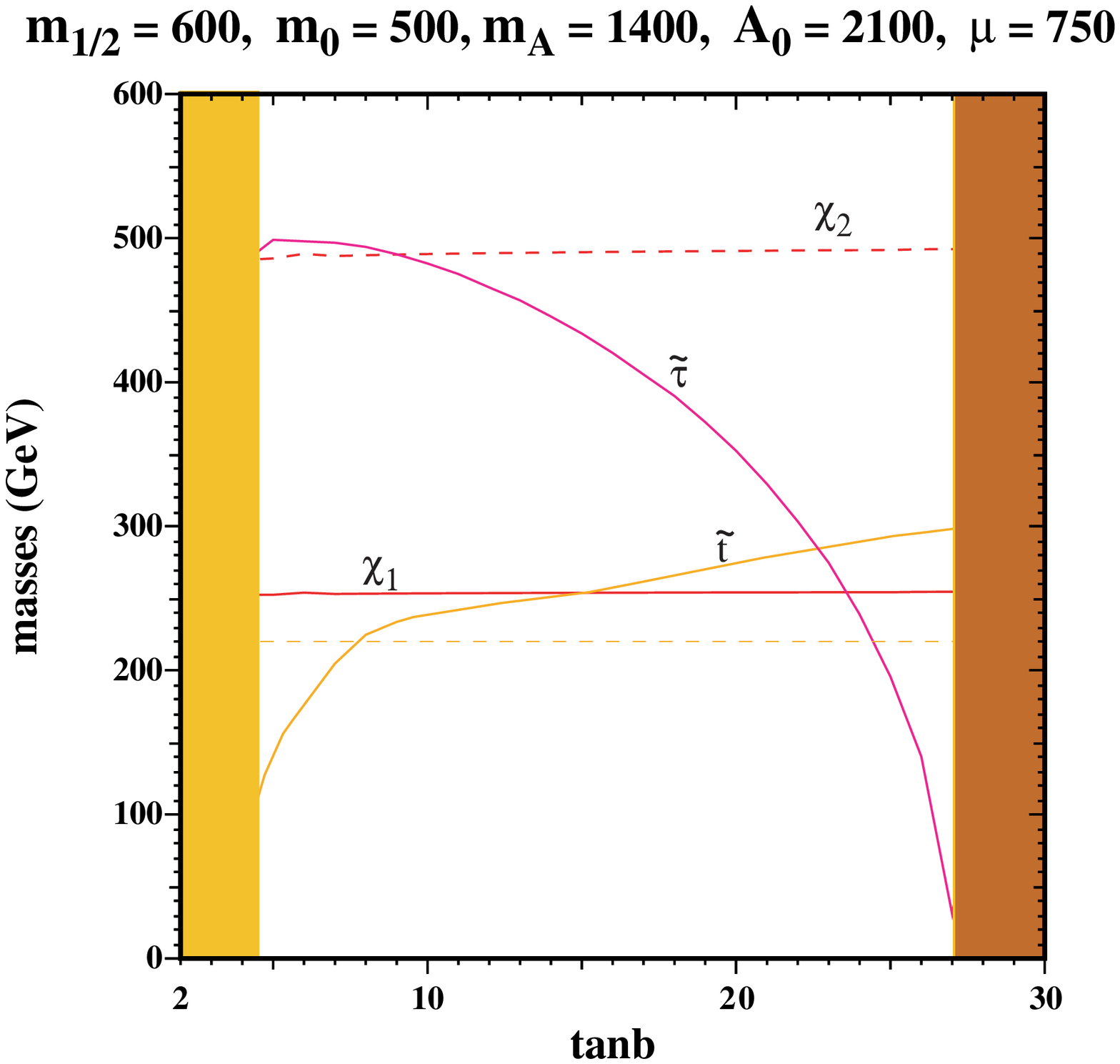,height=7cm}}
\end{center}
\caption{\label{fig:nuhm1b}\it
Sparticle masses in the NUHM as functions of (a) $m_{1/2}$, (b) $m_0$, (c) $A_0$
and (d) $\tan \beta$, with other parameter values specified in the legends. 
}
\end{figure}

In Fig.~\ref{fig:nuhm1b} we show spectra as functions of (a) $m_{1/2}$,
(b) $m_0$, (c) $A_0$ and (d) $\tan \beta$. The lightest neutralino, which is bino-like in the
cases shown, has a mass that depends essentially on $m_{1/2}$ only. We see that $\mstop$
increases as $m_{1/2}$, $m_0$ or $\tan \beta$ increases or $A_0$ decreases,
while $\mstau$ increases as $m_{1/2}$, $A_0$ or $\tan \beta$ decreases or $m_0$
increases. One might attempt to increase $m_{1/2}$ to obtain a heavier NLSP. However,
if we want a stop NLSP, in order to make the stop lighter than neutralino, we need to
compensate the increase in $\mstop$ by (say)
increasing $A_0$, but the constraint on $m_{1/2}$ would in turn render it more difficult to
satisfy the $m_h$ constraint. We see allowed stop NLSP regions in panel (a) for $m_{1/2}
\simeq 595  - 605$~GeV, in panel (b) for $m_0 \simeq 490  - 510$~GeV,
in panel (c) for $A_0 \simeq 2090  - 2110$~GeV, and in panel (d) for $\tan \beta \simeq 8  - 15$.

\begin{figure}
\begin{center}
\mbox{\epsfig{file=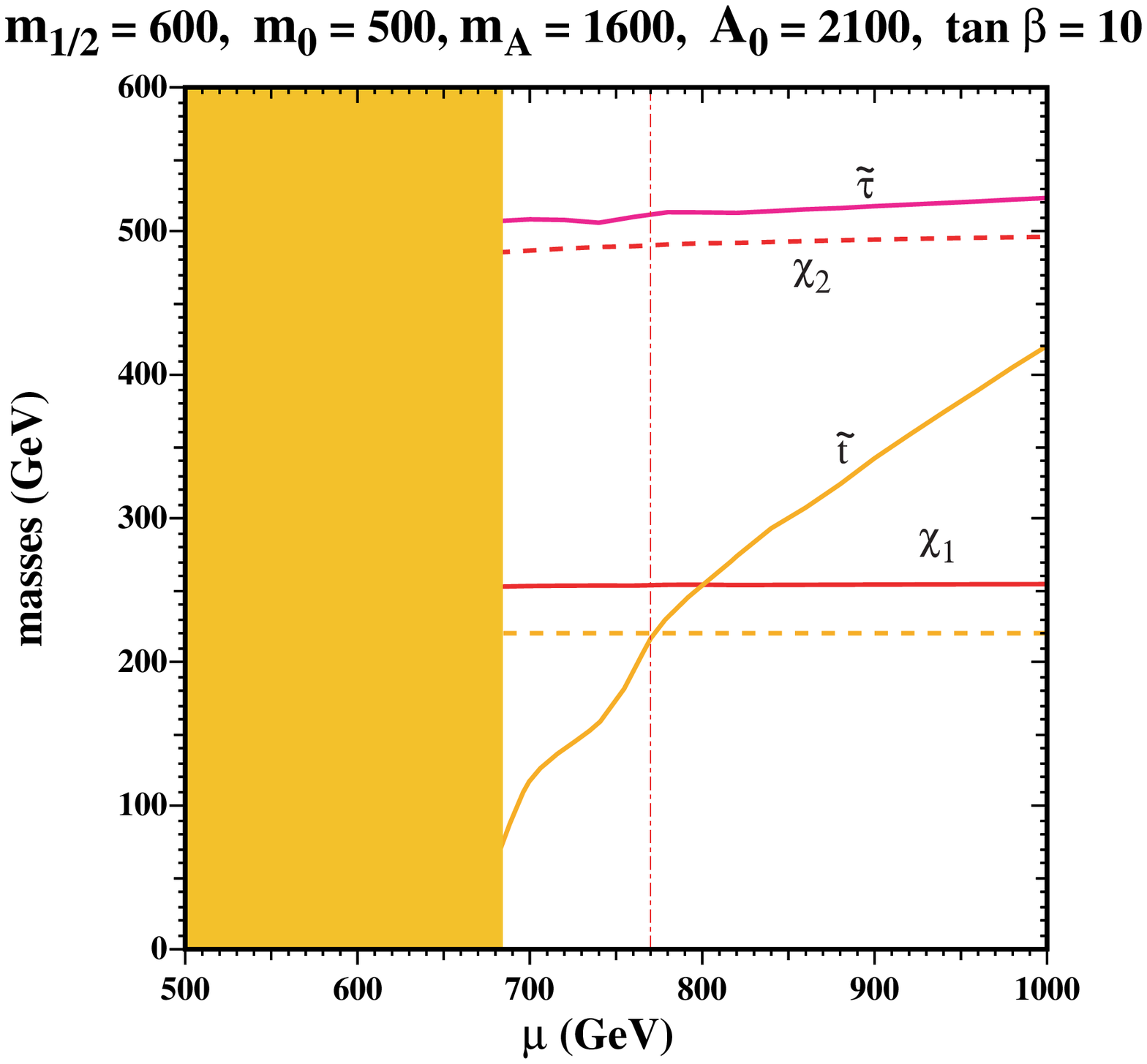,
height=7cm}}
\mbox{\epsfig{file=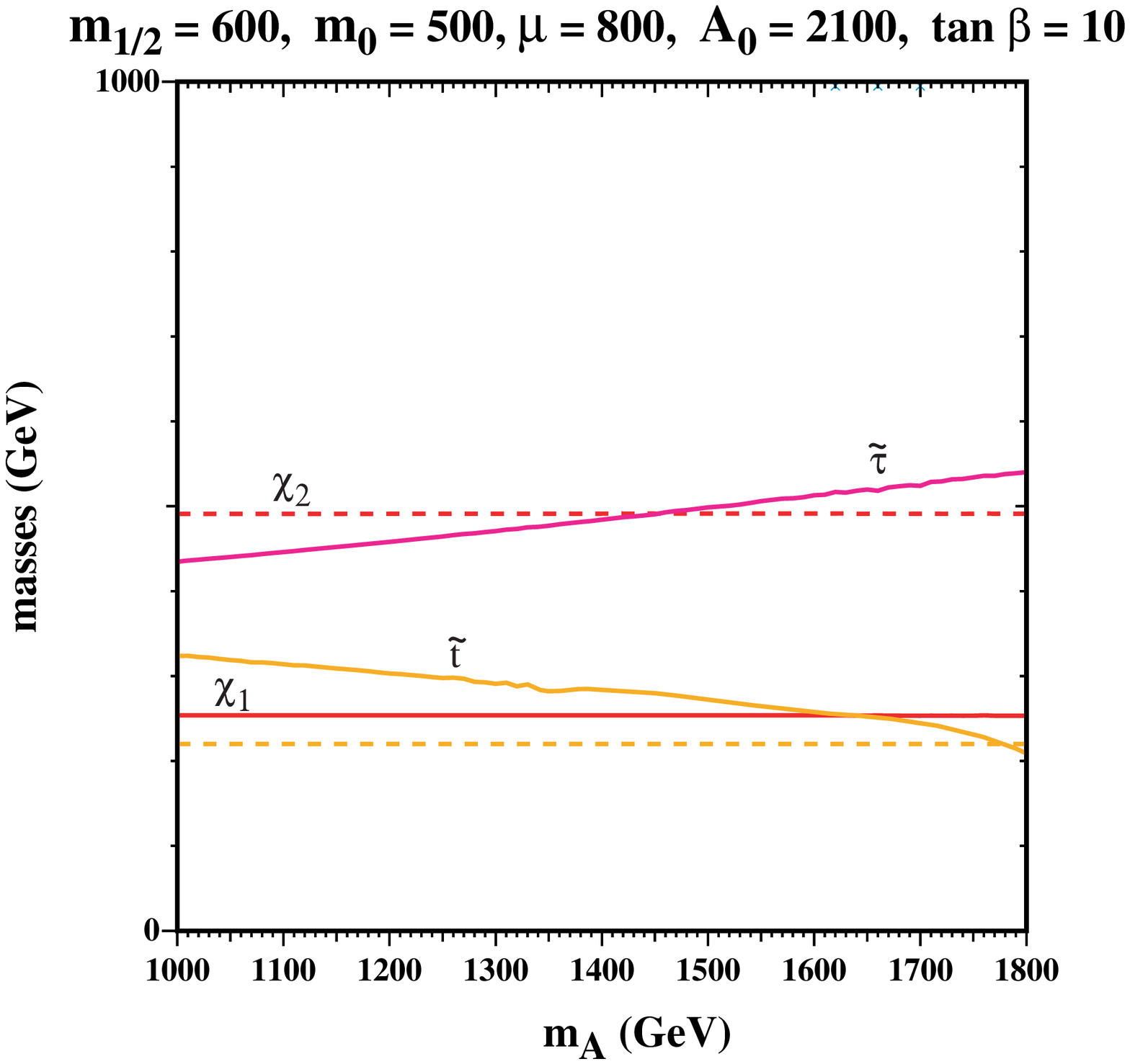,
height=7cm}}
\end{center}
\caption{\label{fig:nuhm2}\it
Same as Fig~\ref{fig:nuhm1}, but with $m_t = 172.7$~GeV.}
\end{figure}

It is well known that the Higgs sector is sensitive to the top
sector. Therefore, it is instructive compare with the result that would hold
if $m_t = 172.7$~GeV (an older experimental value of $m_t$ that is still
within one $\sigma$ of the present central value). We also adjust $m_A$
so as to improve the overlap of the constraints, with the result shown in
Fig.~\ref{fig:nuhm2}.
We see that the Higgs likelihood constraint becomes as strong as the nominal 
$m_h$ taken at face value, which is at $\mu \simeq 770$~GeV, and that
there is only a tiny region allowed. Thus, postulating a larger $m_t$ does not resolve the dilemma of the
Higgs likelihood, even though the face value of $m_h$ is lifted. The allowed region of $\mu$ in
panel (a) of Fig.~\ref{fig:nuhm2} is narrower than in the corresponding panel of Fig.~\ref{fig:nuhm1},
and shifted to larger values of $\mu$. Likewise in panel (b), comparison with
the corresponding panel of Fig.~\ref{fig:nuhm1} shows that the allowed region is smaller
and shifted to larger values of $m_A$.

We conclude that there are some small regions of the NUHM parameter space where the
$\stop_1$ is the NLSP, with a cosmological abundance that would correspond (in the absence
of stop decays) to a relic density $\Omega_{\stop} h^2 \sim 10^{-4}$. Typical allowed
values of the NUHM parameters are $m_{1/2} \sim 600$~GeV, $m_0 \sim 500$~GeV,
$A_0 \sim 2100$~GeV, $\mu \sim 750$~GeV, $m_A \sim 1400$~GeV and
$\tan \beta \sim 10$. We now discuss the
cosmological evolution of such a scenario.

\section{Cosmological Evolution of Metastable Stops}

We expect the metastable stop squarks and antisquarks density to 
have frozen out after coannihilation 
at a temperature $T_F \simeq \mstop/30 \gappeq 7$~GeV~\footnote{Note that the stop
has a larger (co)annihilation cross section than does the neutralino, and hence would have 
smaller freeze out temperature relative to its mass compared to the neutralino
with  freeze-out temperature of $\mchi/20$ to $\mchi/25$.},  
when the age of 
the Universe $t \sim 10^{-9}$ ~s~\footnote{If the stop-neutralino mass difference is
also $\sim 5$~GeV or smaller, which is quite possible in the allowed NUHM regions found in
the previous Section, the stops would have been accompanied by a significant admixture of metastable
neutralinos, which we discuss later.}. One might have expected a primordial stop-antistop
asymmetry comparable to that for conventional baryons, but this would have been eradicated
by stop-stop annihilations. The remnant stops and antistops
would not have decayed before the next major event
in standard Big Bang cosmology, namely the quark-hadron transition when 
$t \sim 10^{-6}$ to $10^{-5}$~s. At this point,  they would have hadronized into sbaryons,
antisbaryons and mesinos. 

Simulations and data from relativistic heavy-ion collisions indicate that the relative 
abundances of hadrons produced at the transition may be modelled by assuming an
effective hadronic freeze-out temperature $T_f \simeq 170$~MeV.
We recall that all the heavier sbaryons
would decay into the lightest $\Lambda_{\widetilde T}$ state, which is charged, whereas the
mesinos would all decay into the lightest ${\widetilde T}^0$ state, which is relatively harmless.
The mass difference between the $\Lambda_b$ baryon and the $B^{\pm,0}$ mesons is
about 365~MeV, and we expect a similar separation for the $\Lambda_{\widetilde T}$ sbaryons 
and the ${\widetilde T}$ mesinos. {\it Modulo} spin and flavour counting factors, this would
suggest a suppression by a factor $\sim 10$ for the abundance of the sbaryons and
antisbaryons relative to the mesinos and antimesinos.
This would suggest {\it a priori} an equivalent relic density 
$\Omega_{\Lambda_{\widetilde T}} h^2 \sim 10^{-5}$, which might have been large enough to
change significantly the subsequent abundances of light elements via bound-state effects.
The $\Xi_{\widetilde T}^+$ state, which is relatively long-lived, would have an abundance about a 
factor of 3 lower than that of $\Lambda_{\widetilde T}$.

It has been argued recently that, following hadronization,
stop hadrons $h_{\widetilde T}$ would capture each other 
and form bound states, leading to the further annihilation of stops~\cite{luty}.
The rate for this process is approximately $n_{h_{\widetilde T}} (T/m_{\ts})^{1/2} m_\pi^{-2}$,
where 
$n_{h_{\widetilde T}} \simeq  \Omega_{h_{\widetilde T}} \rho_c/m_{\ts}$
and
$(T/m_{\ts})^{1/2}$ is the relative velocity in the capture process.
The resulting abundance of stop hadrons is determined by equating this annihilation rate
to the Hubble expansion rate:
\beq
\frac{n_{h_{\widetilde T}}}{m_\pi^2}\left( \frac{T}{m_{\ts}} \right)^{1/2} 
\simeq  \sqrt{\frac{8\pi N}{3}} \frac{T^2}{M_P}.
\eeq
This yields a relic $h_{\widetilde T}$ number density
\beq
\frac{n_{h_{\widetilde T}}}{n_\gamma} \simeq  80 \frac{m_\pi^2}{M_P} 
\frac{m_{\ts}^{1/2}}{T^{3/2}},
\eeq
where we use the estimate $N = {\cal O}(50)$. The $h_{\widetilde T}$ number density
is minimized by annihilations at a temperature close to the formation temperature 
of the bound states, namely $T \sim 200 $ MeV, which yield
\beq
 \frac{n_{h_{\widetilde T}}}{n_\gamma} \simeq 5\times 10^{-17}.
 \eeq
This corresponds to an effective $\Omega_{h_{\widetilde T}} h^2 \sim $ 2 $\times
10^{-6}$ and $\zeta_{h_{\widetilde T}} \equiv n_{h_{\widetilde T}} m_{h_{\widetilde T}} / n_\gamma
\simeq 4 \times 10^{-14}$~GeV for $m_{\tilde t_1} \sim 200$~GeV. With this abundance, the late decays of stop hadrons would probably not cause problems with light-element abundances, even after allowing
for uncertainties in the treatment of showers generated by hadronic decays~\footnote{We note in passing that the mechanism of~\cite{luty} ceases to reduce the ${\widetilde T}$ abundance once
the temperature falls below about 10~MeV, i.e., after about $10^{-3}$~s.}.

However, the catalytic effects of bound states of the negatively-charged metastable relic
antisbaryons ${\bar \Lambda}_{\widetilde T}^-$ are potentially more dangerous, as
emphasized above.
Fortunately, as we now show, their abundance would have been suppressed following hadronization and before BBN, by annihilation with conventional baryons to produce ${\widetilde T}$ antimesinos and
conventional mesons. 
The rate for the annihilation of heavy baryons is approximately
$n_p m_\pi^{-2}$. Since $n_p \gg n_{\Lambda_{\widetilde T}}$ and there is
no velocity suppression for this annihilation process, since $\sigma v \to$
constant as $T \to 0$, the rate for annihilation with
nucleons would be much larger than the capture process~\cite{luty} discussed in the previous
paragraph. The abundance of antisbaryons would remain in equilibrium until the annihilation rate
\beq
\eta\frac{n_\gamma}{m_\pi^2}  \simeq \frac{1.5 \times 10^{-10} T^3}
{m_\pi^2}, 
\label{annrate}
\eeq
where we use $\eta \equiv n_p/n_\gamma = 6.1 \times 10^{-10}$, became of the same order as the expansion rate
\beq
\sqrt{\frac{8\pi}{3}} \frac{(\Omega_m \rho_c)^{1/2}}{M_P} \simeq \frac{1.2 \times 10^{-4} T^{3/2}}{M_P},
\label{exprate}
\eeq
where we use $\Omega_m \sim 0.27$ for the matter density 
as a fraction of the closure density today. The rates (\ref{annrate}) and (\ref{exprate}) become
comparable when
\beq
T \simeq 10^2 \left( \frac {m_\pi^2}{M_P} \right)^{2/3} \simeq 0.1~{\rm eV}.
\label{fout}
\eeq
The $\bar \Lambda_{\widetilde T}^-$ abundance 
would be suppressed by a factor $\exp(-2 \times 10^5)$ already by the time of the onset of BBN at $T \sim 1$~MeV, and the abundance of the 
dangerous $\bar \Lambda_{\widetilde T}^-$ state would be driven to an extremely low
value  $n_{\Lambda_{\widetilde T}} \simeq (m_{\ts} T)^{3/2} e^{-m_{\ts}/T}$ by the time
of freeze-out (\ref{fout}) of annihilations with relic baryons.

The relative abundance of the less dangerous $\Lambda_{\widetilde T}^+$ sbaryons would also 
have been strongly suppressed, as its annihilations with conventional antibaryons
would have continued in equilibrium down to temperatures $T \sim 20$~MeV,
when 
\beq
\frac{(m_p T)^{3/2} e^{-m_p/T}}{m_\pi^2}  \simeq  \sqrt{\frac{8\pi N}{3}} \frac{T^2}{M_P},
\eeq
where $N = {\cal O}(10)$ is the number of relativistic degrees of freedom at $T\sim 20$~MeV.

Thus, during and after BBN, one must contend only with the neutral ${\widetilde T}$ mesinos and
their antiparticles, with which they mix~\cite{GL}. Their decays are relatively innocuous, as 
discussed above. In principle, they might also bind
with deuterium to form superheavy nuclei and thus could change the
Helium abundance.  However, we would not expect the Helium abundance to be
greatly affected: it is at the level of 24\% of the baryon density and hence has 
$\Omega_{He} h^2 \sim 0.01$, whereas by the start of BBN the ${\widetilde T}$ mesino density
would be down at the $10^{-6}$ level in the cases that we
consider here, so the ratio of the number densities (assuming a $\sim 200$~GeV stop
mass) would be ${\cal O}(10^{6})$. We note also that the ${\widetilde T}$ mesinos would
react with conventional baryons to regenerate $\Lambda_{\widetilde T}^+$ sbaryons
after their annihilations with antibaryons had frozen out, i.e., at $T < 20$~MeV. However, this
process could generate a $\Lambda_{\widetilde T}^+$
density of at most $10^{-6}$, which would not be problematic.

This analysis therefore finds no cosmological problems with 
most of the small region of NUHM parameter 
space discussed in Section~6 where the stop is the NLSP.

\section{On the Possibility of a Metastable Neutralino}

The conclusion of the previous Section may, however, be modified if the
stop NLSP is nearly degenerate with the lightest neutralino $\chi$, a possibility
suggested by our analysis of the NUHM parameter space. The $\chi$ may decay either into the stop (or
antistop) and some other
particles: $\chi \to {\tilde t_1}+ X, \ {\tilde
t_1}^\ast + X^\prime$, or into ${\widetilde G} + \gamma$. The latter decay is very slow, being of
gravitational strength~\cite{gdm}:
\beq
\Gamma (\chi \to \widetilde{G} \gamma) = \frac{1}{16 \pi} \frac{C_{\chi
\gamma}^2}{M_P^2} \frac{m_\chi^5}{m_{\widetilde G}^2} \left( 1 -
\frac{m_{\widetilde G}^2}{m_\chi^2} \right)^3 \left( \frac{1}{3} +
\frac{m_{\widetilde G}^2}{m_\chi^2} \right)  ,
\eeq
where $C_{\chi \gamma} = (O_{1\chi} \cos \theta_W + O_{2 \chi} \sin \theta_W)$.
The decay rate depends on $m_{\widetilde G}$, and for $m_{\widetilde G} \simeq 1 -
200$~GeV we get $\Gamma (\chi \to \widetilde{G} \gamma) \simeq 10^{-28} -
10^{-34}$~GeV corresponding to lifetime of $\sim 10^4 - 10^{10}$~s.
If the lightest neutralino is almost 
degenerate with the stop, the two-body decays $\chi \to \stop_1 +
\bar{t}, \ {\bar \stop_1} + t$, 
which would be rapid but require a large mass difference $m_\chi - \mstop > m_t$, 
are not available. As we discuss below, the decays $\chi \to {\tilde t_1}+ X, \ {\tilde
t_1}^\ast + X^\prime$ may be very
suppressed if the $\chi$ and $\stop_1$ are sufficiently degenerate.
If they are suppressed sufficiently for the $\chi$ lifetime 
to exceed about $10^{-5}$~s, neutralino 
relics would be present during and after hadronization, and the cosmology of the stops
produced in their decays would differ from that of the thermally-produced stops discussed in
the previous Section, as we show below.

To show that the abundance of neutralino relics could indeed be significant,
recall that when the stop and neutralino are almost degenerate, i.e., $m_\chi - m_{\stop_1}$ less
than the freeze-out temperature $\sim m_{\stop_1} / 30$, neutralinos are kept in thermal equilibrium 
through co-annihilation processes \cite{GS} such as $\chi X \leftrightarrow \stop_1 X^\prime$
where $X,X^\prime$ are Standard Model particles. We can approximate the neutralino relic 
density between the epochs of supersymmetric freeze-out and relic decays by
\beq
\Omega_\chi \simeq \frac{1}{3} \ \Omega_{\stop} \simeq 0.25 \, \Omega_{NLSP} ,
\eeq
modulo the Boltzmann factor from the mass difference. Here the factor 3 is due
to colour, and  $\Omega_{NLSP} = \Omega_\chi + \Omega_{\stop}$  
is the relic density one obtains
from a standard coannihilation calculation, which leads for near-degenerate $\chi$ and $\stop_1$ to 
$\Omega_{NLSP} h^2 \sim 1.6 \times 10^{-4}$ and hence $\Omega_\chi h^2 \sim 4 \times 10^{-5}$. 
Therefore the novel cosmology of the stops produced in
neutralino decays is potentially important.

If the $\chi - \stop_1$ mass difference is larger than about 1~GeV, 
the neutralino may decay into three bodies, e.g., $\chi \to \stop_1 + {\bar s} +
\pi^-$, and into four bodies, e.g.,
$\chi \to \stop_1 + {\bar s} + \ell + \nu$ and conjugate modes, and also into the
corresponding final states with ${\bar s} \to {\bar d}$ and/or $\ell + \nu \to q + {\bar q}$.
By analogy with the result of~\cite{jittoh}, the four-body semileptonic decay rate
can be approximated as:
\begin{equation}
\Gamma (\chi \to \stop_1 {\bar s} \ell \nu) = {\cal O}(\frac{1}{100}) \frac{G_F^2 |V_{ts}|^2}{(2\pi)^5 \mstop m_t^4} 
\times
\left[ 4 |g_L|^2 (\delta m)^{10} - 8 \, {\rm Re}[g_L^\ast g_R] m_t (\delta m)^9 + 6 |g_R|^2 m_t^2 
(\delta m)^8) \right]  ,
\label{4chi}
\end{equation}
where we allow for both $\stop_1$ and $\stop_1^\ast$ modes and a factor of 3
for colour. We have neglected the final-state fermion masses and use 
\bear
g_L &=& - \frac{g}{\sqrt{2}} \left\{ \cos \theta_t \left[ O_{\chi 2} +
\frac{\sin \theta_W}{3 \cos \theta_W} O_{\chi 1} \right] + \sin \theta_t
\frac{m_t}{m_W \sin \beta} O_{\chi 4} \right\}, \\
g_R &=& - \frac{g}{\sqrt{2}} \left\{ \cos \theta_t \frac{m_t}{m_W \sin \beta}
O_{\chi 4} - \sin \theta_t \frac{4 \sin \theta_W}{3 \cos \theta_W} O_{\chi 1}
\right\} ,
\eear
where $O$ is the neutralino diagonalization matrix $O^T M_N O = M_N^{diag}$.
We see that the last term in
the square bracket in (\ref{4chi}) is dominant, due to the large magnitude of $m_t$.
Therefore, we can approximate further to obtain
\beq
\Gamma (\chi \to \stop_1 {\bar s} \ell \nu) \sim {\cal O}(\frac{1}{25}){\frac{G_F^2 |V_{ts}|^2 |g_R|^2
(\delta m)^8}{(2\pi)^5 \mstop m_t^2}} .
\eeq
For $\mstop = 220$~GeV and $\delta m = 1$~GeV, for example, and including the 
$e \nu, \mu \nu$ and $d {\bar u}$ final states, we estimate $\Gamma
(\chi \to \stop_1 {\bar s} f {\bar f}) \sim {\cal O}(10^{-26})$~GeV, i.e., a $\chi$
decay lifetime $\sim 10^{2}$~s. 

The partial decay time would be even longer
for yet smaller $\delta m$, but it would be necessary to take into account bound-state
effects in the $\stop_1 {\bar q}$ and $q {\bar q}$ channels, and even in the full
four-body final state. A naive scaling by $(\delta m)^8$ would yield a
lifetime in excess of $10^{10}$~s for $\delta m \sim 0.1$~GeV. It is therefore possible
that the partial lifetime for $\chi \to \stop_1 + X, {\tilde t_1}^\ast + {\bar X}$ decays could 
exceed the partial lifetime for $\chi \to {\widetilde G} + \gamma$ decays.

As already remarked,
if the $\chi$ lifetime exceeds about $10^{-5}$~s, the $\stop_1$ decay products
appear after the quark-hadron transition, and the discussion of stop cosmology
given in the previous Section must be modified. If $\tau_\chi$ exceeds about
$10^{-3}$~s, the mechanism of~\cite{luty} becomes ineffective for reducing the abundance
of stop hadrons produced in $\chi$ decays, which therefore remain comparable to
$\Omega_\chi h^2 \sim 4 \times 10^{-5}$, and also annihilations with antiprotons become ineffective
for reducing the $\Lambda_{\widetilde T}^+$ abundance. On the other hand,
annihilations with baryons remain effective for converting the 
dangerous $\Lambda_{\widetilde T}^-$ sbaryons into relatively innocuous
${\widetilde T}^0$ mesinos for any $\chi$ lifetime up to about $10^{14}$~s. If $\tau_\chi
< \tau_{\stop_1}$, the BBN/CMB constraints on 
electromagnetic and hadronic stop decays are insensitive to $\tau_\chi$, and depend only
on $\tau_{\stop_1}$. If $\tau_\chi > \tau_{\stop_1}$,
the the BBN/CMB constraints on electromagnetic and hadronic stop decays should be
evaluated with $\tau_{\stop_1}$ replaced by $\tau_\chi$. Finally, if the partial lifetime for 
$\chi \to \stop_1 + X, {\tilde t_1}^\ast + {\bar X}$ decays exceeds that for $\chi \to 
{\widetilde G} + \gamma$ decays, the $\chi$ decay contribution to the relic stop density is
diluted by the ratio of the partial decay rates.

We therefore distinguish five metastable neutralino cases.

\begin{enumerate}
{\item If $\tau_\chi < 10^{-3}$~s, the residual suppression of the ${\widetilde T}^0$ mesino
density the mechanism of~\cite{luty} may still be sufficient to evade the BBN/CMB constraints 
on electromagnetic and hadronic stop decays.}
{\item If  $10^{-3}$~s $< \tau_\chi < 10^4$~s, the the mechanism of~\cite{luty}
is ineffective, and the indirectly-produced ${\widetilde T}^0$ mesinos still have
$\Omega_\chi h^2 \sim 4 \times 10^{-5}$. Since the neutralinos decay before the epoch when
the light-element abundances constrain the electromagnetic and hadronic stop decays, these
depend on the value of $\tau_{\stop_1}$ in the usual way. The relic density of ${\widetilde T}^0$
mesinos respects the limit obtained in~\cite{CEFO} by considering electromagnetic decays,
but not the hadronic limit given in~\cite{othercosm}, which will be re-evaluated in~\cite{CEFOS2}.}
{\item The usual light-element abundance constraints also apply if $10^4$~s $< \tau_\chi < 
\tau_{\stop_1}$, and the relic density of ${\widetilde T}^0$ mesinos is again
marginal.}
{\item If $\tau_\chi > \tau_{\stop_1}$ and $10^4$~s
simultaneously, the usual light-element abundance constraints
still apply, but they should be evaluated with $\tau_{\stop_1}$ replaced by
$\tau_\chi + \tau_{\stop_1}$. In addition, we should consider also the effect of
neutralino decay itself to the light element abundance. }
{\item In both the last two cases, the scenario may survive more easily if the partial lifetime for 
$\chi \to \stop_1 + X, {\tilde t_1}^\ast + {\bar X}$ decays exceeds that for $\chi \to 
{\widetilde G} + \gamma$ decays, in which case the abundance of
${\widetilde T}^0$ mesinos is suppressed by the ratio of the partial decay rates.}
\end{enumerate}

\section{Summary}

The scenario of gravitino dark matter with a stop NLSP has rich phenomenology,
for both collider experiments and cosmology.
Unfortunately this scenario is disfavoured in the CMSSM and NUHM, due to the
existing collider limits and the close relation between the masses of the light stop and 
the light Higgs boson. However, we find that there could still be a small allowed region, at least within the NUHM, with $m_{1/2} \sim 600$~GeV, $m_0 \sim 500$~GeV,
$A_0 \sim 2100$~GeV, $\mu \sim 750$~GeV, $m_A \sim 1400$~GeV and
$\tan \beta \sim 10$.

Much of the discussion in this paper on the stop NLSP scenario could be applied as well to 
a scharm or sup NLSP in more general MSSM models with a
gravitino LSP. In such a case, we expect that the Higgs constraint would be
less stringent.  Presumably the late production of the NLSP would be less important
since, with the light $u,c$ quarks in place of the heavy $t$ quark, two-body decay channels
would probably be available, unless the mass difference between neutralino and
the NLSP is very small (less than $m_{u,c}$), in
which case the neutralino would decay directly into the gravitino.
The case of a sbottom NLSP would in this respect be intermediate between the stop
and scharm cases. However, as noted above, the lightest mesino would, in this case,
probably be charged, and hence capable of bound-state catalysis of dangerous light-element
transmutations.

For a possible realization of models with a sup NLSP, see~\cite{baerNUHM},
where one could get light scharm and sup in the NUHM by assuming very high values
for $\mu$ and $m_A$, which would mean violating the GUT stability constraint. Another
possiblity would be to postulate non-universality between the first two generations and
the third, and between squarks and sleptons. It is interesting to note that the
suppression of the lighter squark eigenmass via diagonalization is much less for sup than 
for stop, so it is not necessary to postulate large $A_0$ in these scenarios.  

As Hamlet said: `There are more things in Heaven and Earth, Horatio, than are dreamt of
in your philosophy.' This comment certainly applies to supersymmetric phenomenology.
There are surely many important aspects of the stop NLSP scenario that we have overlooked
in this paper, and many other NLSP candidates could be envisaged, beyond the stop, the
neutralino and the stau, which are the three options usually considered. If supersymmetry
is the `surprise' most expected at the LHC, one should perhaps expect it to appear in
an unexpected way. The dominant signature might not be the `expected' missing transverse
energy, but rather some brand of metastable charged particle, which could well have strong
interactions, like the stop considered here.

\section*{Acknowledgments}
\noindent 
The work of K.A.O. was supported in part
by DOE grant DE--FG02--94ER--40823. 
The work of Y.S. was supported in part
by the NSERC of Canada. 
J.L. D.-C. thanks SNI and CONACYT (Mexico) for their support, and the
HELEN project for financing a visit to CERN. JE is pleased to thank
Oleg Lebedev for useful discussions,
and YS is pleased to thank Maxim Pospelov and Rich Cyburt for useful discussions.
We thank Teruki Kamon for information on the CDF search on (meta)stable charged
massive particles.

\section*{Appendix: Three-Body Stop Decays}

\noindent
We present in this Appendix some details of the calculation of the three-body stop decay, using
the following notations. For the top contribution, we introduce:
\begin{eqnarray}
A_{\ts} & \equiv & \frac{1}{2} ( \cos\theta_{\ts}+\sin\theta_{\ts} ), \\
B_{\ts} & \equiv & \frac{1}{2} ( \cos\theta_{\ts}-\sin\theta_{\ts} ).
\end{eqnarray}
and for the sbottom contribution we introduce:
\begin{eqnarray}
a_i & \equiv & ( \sin \theta_{\bs},\cos\theta_{\bs} ), \\
b_i & \equiv & (\cos\theta_{\bs},-\sin\theta_{\bs}).
\end{eqnarray}
We also define
\beq
\kappa_i \equiv (\cos\theta_{\ts} \cos\theta_{\bs} ,- \cos\theta_{\ts}
\cos\theta_{\bs} ).
\eeq
In the chargino contribution we have:
\begin{eqnarray}
V_i & \equiv & \frac{1}{2} ( V_{i2} \sin\beta + U_{i2} \cos\beta ), \\
A_i & \equiv & \frac{1}{2} (V_{i2} \sin\beta - U_{i2}\cos\beta),
\end{eqnarray}
and for the low-to-moderate range of $\tan\beta$ we have:
\beqn
2S_1&=& -g_2 \cos\phi_L+ \frac{g_2 m_t \sin \phi_L \sin\theta_{\ts}}{ \sqrt{2} m_W \sin\beta} \\
2P_1&=& -g_2 \cos\phi_L- \frac{g_2 m_t \sin \phi_L \sin\theta_{\ts}}{ \sqrt{2} m_W \sin\beta}
\eeqn
where $\cos \phi_L, \pm \sin \phi_L$ are elements of the matrix $V$ that diagonalizes
the chargino mass matrix, and expressions for $S_2$ and $P_2$ may
be obtained  by replacing $\cos\phi_L \to - \sin\phi_L$ and $\sin\phi_L \to  \cos\phi_L$
in the last equations. 

For the {\bf {\it square of the top contribution}} to the decay amplitude, we have:
\beq
W_{tt}= \frac{4}{m^2_W \mgrav^2} h_1 [ h_2 ( (A_{\ts}+B_{\ts})^2 m^2_t + (A_{\ts}-B_{\ts})^2 q^2_1) 
                                  + h_3 ( (A_{\ts}^2-B_{\ts}^2)\mgrav m_t +
				  (A_{\ts}-B_{\ts})^2 (f_2-\mgrav^2) )] ,
\eeq
where
\bea
h_1&=&  \mstop^2 \mgrav^2 - f^2_2 , \\
h_2&=& -2 f_1 f_3+ 2 f^2_3+m^2_W(-f_2+3 f_3+\mgrav^2) , \\
h_3&=& -2m^2_W q^2_1 -2(f_1-f_3)(2f_1-2f_3- 3 m^2_W) . 
\eea
From the energies $E_1=p^0_1$ and $E_W=k^0$, we can define variables $x,y$ by
$E_1= m^2_{{\ts}_1} x/ 2$ and
$E_W= m^2_{{\ts}_1} y/ 2$. In turn, this allows to express all the inner
products of momenta that appear in the functions $W_{ii}$ in terms of
$x$ and $y$. Then, $q^2_1=\mstop^2(1+r_1-x)$ and the  functions $f_i$ are given as follows:
\bea
f_1&=& \frac{\mstop^2}{2} y  ,  \\
f_2&=&  \frac{\mstop^2}{2} x , \\
f_3&=& \frac{\mstop^2}{2} (-1-r_1-r_2+x+y) . 
\eea
where $r_1=\mgrav^2/\mstop^2$,  $r_2=m^2_W/\mstop^2$.

Similar expressions can be written for the {\bf {\it square of the sbottom contribution}}
to the decay amplitude, namely:
\beq
{|M_{\sbot}|}^2 = \sum_{i,j} C^\ast_{\sbot_i} C_{\sbot_j} P^\ast_{\sbot_i}(q_2)
P_{\sbot_j}(q_2) W_{\sbot_i \sbot_j}
\eeq
where:
\beq
W_{\sbot_i \sbot_j}= \frac{16}{3} \frac{a_i a_j + b_i b_j}{m^2_W \mgrav^2} (\mstop^2
m^2_W - f^2_1)( q^2_2 \mgrav^2 -(q_2.p_1)^2 )\, p_1.p_2
\eeq
and: $q^2_2= \mstop^2 (1+r_2-y)$, $q_2.p_1= f_2-f_3$ and $p_1.p_2=
f_2-f_3-\mgrav^2$.

The {\bf {\it square of the chargino contribution}} takes the form:
\beq
|M_{\chi^+}|^2 = \sum_{i,j} C^2_{\chi^+} P^\ast_{\chi^+_i}(q_3) P_{\chi^+_j}(q_3)
W_{\chi^+_i \chi^+_j}
\eeq
where
\beq
W_{\chi^+_i \chi^+_j}= \frac{32}{3 m^2_W \mgrav^2} (2m^2_W \mgrav^2 + f^2_3) 
          \left[ 2 f_4 ( \Sigma_{(1)ij} \mgrav  m_{\chi} 
	  + \Sigma_{(2)ij} f_5)
	  + p_1.p_2 \, ( m^2_{\chi}
	  \Sigma_{(3)ij}- q^2_3 \Sigma_{(2)ij} ) \right]
\eeq                       
and $q^2_3= \mgrav^2 + m^2_W+2 f_3$, $f_4= (\mstop^2/2) (2-x-y)$, $f_5=
f_3+\mgrav^2$.
Also, we define $\Sigma_{(1)ij} \equiv (A_i A_j - V_i V_j)(S_iS_j+P_i P_j)$, 
$\Sigma_{(2)ij} \equiv (A_i A_j + V_i V_j)(S_iS_j+P_i P_j) - (A_i V_j +
V_i A_j)(S_i P_j + P_i S_j)$,
and   $\Sigma_{(3)ij} = (A_i A_j + V_i V_j)(S_iS_j+P_i P_j) + (A_i V_j + V_i
A_j)(S_i P_j + P_i S_j)$.

The {\bf {\it interference between the top and chargino contributions}} leads to the
following expression for $W_{t\chi^+}$:
\beq
W_{t\chi^+_i}= 2(T_{1i}-m_t T_{2i})+2 m_{\chi^+_i}(T_{3i}-m_t T_{4i})-
\frac{2}{3} (T_{5i}-m_t T_{6i}) - 
                       \frac{2 m_{\chi^+_i} }{3} (T_{7i}-m_t T_{8i})
\eeq
The functions $T_i$ may be written as follows:
\beqn
T_{1i} &=& \frac{4 \, \alpha_{1i}}{\mgrav m^2_W} [ -2f^3_1 \mgrav^2 + (f_2 f^2_3-
\mgrav^2 f_1 f_3) h_4 
         - 2f_1 f_2 f_3(-f_2+2 f_3+ \mgrav^2)  \nonumber \\
    & & - 2f^3_2 m^2_W + f_2 m^2_W (f^2_3+ 2 \mgrav^2 m^2_{{\ts}_1})  
     - f_1 m^2_W ( 2 f^2_2+f_2f_3+ \mgrav^2 f_3 - 2 \mgrav^2 m^2_{{\ts}_1}) \nonumber \\
    & & + m^2_W( f^2_2 - \mgrav^2 m^2_{{\ts}_1} ) h_5 
        + f^2_1 ( 2 f_2f_3 - 2 \mgrav^2 f_2+ \mgrav^2(4f_3+2\mgrav^2+m^2_W) ) ] , 
\\
T_{2i} &=&  \frac{4 \, \alpha_{2i}}{\mgrav^2 m^2_W} [ (f_1-f_3)(2f_3+\mgrav^2)(\mgrav^2
f_1-f_2f_3) \nonumber \\
    & & + m^2_W ( -f_2 f^2_3 -f^2_2(f_3+ \mgrav^2) + \mgrav^2( f_1
    f_3+f_3\mstop^2 + \mgrav^2 \mstop^2) ) ] ,  
\\
T_{3i}&=& \frac{4 \, \alpha_{3i}}{\mgrav^2 m^2_W}[ 2f^2_1 \mgrav^2 (  \mgrav^2-f_2)+ 
          f_2f^2_3( \mgrav^2 - m^2_{{\ts}_1})                          \nonumber \\
   & &   + f_1 f_3( 2f^2_2-2 f_2 \mgrav^2 - \mgrav^4+ \mgrav^2 m^2_{{\ts}_1}) \nonumber  \\
   & &  - m^2_W  (f^2_2 - \mgrav^2 m^2_{{\ts}_1})(2f_2 -f_3-2 \mgrav^2)] , 
\eeqn
\beqn
T_{4i}&=& -\frac{4 \, \alpha_{4i}}{\mgrav m^2_W}  [ (f_1-f_3)(\mgrav^2 f_1-f_2 f_3)+
m^2_W (f^2_2- \mgrav^2 m^2_{{\ts}_1} )] ,
\\
T_{5i}&=& \frac{4 \, \alpha_{1i}}{\mgrav m^2_W} [-2 f^2_1(f_2- \mgrav^2)(f_3+ \mgrav^2)
+ f^2_3f_2( 2f_3+ \mgrav^2)
                - f^2_3 m^2_{{\ts}_1} (f_2+2f_3) \nonumber \\
   & & +f_1f_3 (2 f^2_2-2 \mgrav^2 f_2-(2f_3+ \mgrav^2)( \mgrav^2 -
		m^2_{{\ts}_1}) ) \nonumber  \\
   & & + m^2_W f_2 ( - 2 f^2_2+5f_2 f_3+ f^2_3+ 2 \mgrav^2 f_2)\nonumber \\
   & & - m^2_W m^2_{{\ts}_1} ( 2f_2 (f_3- \mgrav^2) + (f_3+ \mgrav^2)(f_3+ 2
   \mgrav^2) ) \nonumber \\
   & & m^2_W f_1 ( f_2 (f_3-2 \mgrav^2)- \mgrav^2 (f_3-2  m^2_{{\ts}_1})) + 2
   m^4_W (f^2_2 - \mgrav^2 m^2_{{\ts}_1} )] ,
\\
T_{6i}&=& \frac{4 \, \alpha_{2i}}{ \mgrav^2 m^2_W}[ -2 f^2_1 \mgrav^2 (f_3+ \mgrav^2) +
m^2_W \mgrav^2 f_1 (2f_2+f_3) \nonumber  \\
   & & + f_1 f_3( 2  f_2 f_3 + \mgrav^2 (3f_2+2f_3)+ \mgrav^4 )\nonumber  \\
   & & + f_3 ( -m^2_W f^2_2- \mgrav^2 m^2_{{\ts}_1}(f_3+m^2_W) - f_2 f_3(2f_3+
   \mgrav^2 +m^2_W)) ]  , 
\\
T_{7i}&=& \frac{4 \, \alpha_{3i}}{\mgrav^2 m^2_W} [ f^2_1 (-2 \mgrav^2 f_2+2 \mgrav^4)+
f_2 f^2_3 ( \mgrav^2 - m^2_{{\ts}_1})  \nonumber \\
   & & + f_1 f_3 ( 2f^2_2- 2 f_2 \mgrav^2 - \mgrav^4+ \mgrav^2 m^2_{{\ts}_1} ) \nonumber , \\
   & & - f_2 m^2_W(  f_2 (2f_2-f_3)- 2 \mgrav^2 (f_2+f_3) ) - \mgrav^2
   m^2_{{\ts}_1} m^2_W (-2f_2+3f_3+2 \mgrav^2) ] ,
\\
T_{8i}&=& \frac{4 \, \alpha_{4i}}{\mgrav m^2_W} [ 2 \mgrav^2 f^2_1+ f_3( f_2f_3 + f_3 m^2_{{\ts}_1} + 2f_2 m^2_W ) 
    -f_1( 3f_2 f_3 + \mgrav^2 ( f_3 + 2 m^2_W ) ) ],
\eeqn
where
$\alpha_{1i}=\Delta_{3i} A_{\ts} + \Delta_{4i} B_{\ts}$,
$\alpha_{2i}=\Delta_{3i} A_{\ts} - \Delta_{4i} B_{\ts}$, 
$\alpha_{3i} =\Sigma_{3i} A_{\ts} + \Sigma_{4i} B_{\ts}$, 
$\alpha_{4i} =\Sigma_{3i} A_{\ts} - \Sigma_{4i} B_{\ts}$, 
with  $\Delta_{3i}= V_i S_i -A_i P_i$, $\Sigma_{3i}= V_i S_i + A_i P_i$ and 
 $\Delta_{4i}= V_i P_i -A_i S_i$, $\Sigma_{4i}= V_i P_i + A_i S_i$. 
 We also denote
$h_4= 2f_3+ \mgrav^2 - m^2_{{\ts}_1} $, $h_5= 3 f_3  + 2 \mgrav^2 +m^2_W $.

The {\bf {\it interference between the top and sbottom contributions}}, $W_{t\bs}$ 
may be written as:
\beq
W_{t\bs_i}= \frac{ (f_2 \, q_2.q_1 - p.q_2 \mgrav^2) } {\mgrav^2 m^2_W} [ m^2_W
Y_{1pi}- f_1 Y_{1ki} ]
                                + \frac{1}{3m^2_W} [ m^2_W Y_{2pi}- f_1 Y_{2ki}] ,
\eeq
where
\beqn
Y_{1pi}&=& 4 \, Z_{1i} [ (f_1+f_2) \mgrav^2 + (f_2 - f_3 - 2 \mgrav^2) m^2_{{\ts}_1}] \nonumber \\
      & &    - 4 \, Z_{2i} [ (f_1+f_2- m^2_{{\ts}_1}) m_t \mgrav] ,
\\
Y_{1ki}&=& 4 \, Z_{1i} [ 2f_1 (f_2- \mgrav^2) + f_3 ( \mgrav^2 -  m^2_{{\ts}_1}) +m^2_W
(-f_2+ \mgrav^2)] 
\nonumber  \\
      & &    - 4 \, Z_{2i} [ (-f_1+f_3- m^2_W) m_t \mgrav] ,\\
Y_{2pi}&=& \frac{ 4 \, Z_{1i}}{\mgrav^2} [ f^3_3( \mgrav^2 + m^2_{{\ts}_1})+ 
               f_2 m^2_{{\ts}_1} (f^2_3+2f_3 \mgrav^2 -\mgrav^2 ( \mgrav^2 + m^2_{{\ts}_1}) ) \nonumber \\
 & & + \mgrav^2 m^2_{{\ts}_1} ( - f^2_3+ f_3 m^2_{{\ts}_1}+ \mgrav^2 (
	       2 m^2_{{\ts}_1} + m^2_W) ) \nonumber \\
      & & - f^2_2 ( f_3 (\mgrav^2 + 2 m^2_{{\ts}_1}) + \mgrav^2 ( 2 m^2_{{\ts}_1}+ m^2_W ) ) \nonumber \\
      & & +\mgrav^2 f_1 ( f_2 (f_2 +f_3 + \mgrav^2)+ m^2_{{\ts}_1} (f_2 - f_3 -3
      \mgrav^2) ) ] \nonumber \\
      & & + \frac{ 4 \, Z_{2i}}{\mgrav^2} [  \mgrav m_t (- f^3_3 - f_2 m^2_{{\ts}_1} 
      (f_3- \mgrav^2)  \nonumber  \\
      & & +  f^2_2 ( f_3 + m^2_{{\ts}_1}+ m^2_W ) +  m^2_{{\ts}_1} ( f^2_3 -
      \mgrav^2 ( m^2_{{\ts}_1}+ m^2_W ) ) \nonumber \\
      & &           +  \mgrav m_t f_1 ( - f_2 (f_2+f_3+ \mgrav^2)+ 2 \mgrav^2 m^2_{{\ts}_1} ) ] , 
\\
Y_{2ki}&=& \frac{ 4 \, Z_{1i}}{\mgrav^2} [ ( f^2_2 -f_2 f_3- \mgrav^2 m^2_{{\ts}_1} ) 
                              ( f_3 ( - \mgrav^2 + m^2_{{\ts}_1} ) + m^2_W (f_2-
			      \mgrav^2) ) \nonumber \\
      & & -f_1 (  2f^3_2 - 2f^2_2 (f_3+ \mgrav^2) + \mgrav^2 f_3( \mgrav^2- m^2_{{\ts}_1})\nonumber  \\
      & &        + \mgrav^2 f_2 (2f_3 - 2 m^2_{{\ts}_1} -m^2_W) + \mgrav^4 ( 2 m^2_{{\ts}_1} + m^2_W ) )\nonumber  \\
      & & -2 \mgrav^2 f^2_1 (f_2- \mgrav^2) ] \nonumber \\
      & & + \frac{ 4 \, Z_{2i}}{\mgrav^2} [ -f_3 \mgrav m_t ( -f^2_2 + (f_3+ \mgrav^2) m^2_{{\ts}_1} + f_2(f_3+m^2_W))\nonumber  \\
      & &     + f_1 \mgrav m_t ( -f^2_2 + 3f_2 f_3 + \mgrav^2 (f_3+
      m^2_{{\ts}_1}+m^2_W ) -2 f_1 \mgrav^2 ) ],
\eeqn
and $Z_{1i}=A_{\ts} \Sigma_{5i} + B_{\ts} \Delta_{5i}$, $Z_{2i}=A_{\ts}
\Sigma_{5i} - B_{\ts} \Delta_{5i}$, $\Sigma_{5i}=(a_i+b_i)/2$, and
$\Delta_{5i}=(a_i-b_i)/2$.

Finally, the {\bf {\it interference between the chargino and sbottom contributions}} is given by:
\beq
W_{\chi^+_i \bs_j}= 2[X_{1pij}- m_{\chi^+_i} X_{2pij} ]
                                - \frac{2}{3} [ X_{1kij}- m_{\chi^+_i}  X_{2kij} ]
\eeq
where
\beqn
X_{1pij}&=& -\frac{4 \, \sigma_{1ij}}{\mgrav m^2_W} [ (-f_1-f_2+2f_3+ \mgrav^2+m^2_W)( 
h_6  + h_7 - \mgrav^2 f^2_1) ]  ,\\
X_{1kij}&=& -\frac{4\, \sigma_{1ij}}{\mgrav m^2_W} [ f^2_1 ( f_3 ( f_2+f_3 ) + \mgrav^2
(f_2 -2 f_3)- \mgrav^4  ) \nonumber \\
      & &   -f_1 f_3( f^2_2 - f_2( 3f_3+ \mgrav^2 -2m^2_W ) + f_3 ( 2f_3 +
      \mgrav^2 +  m^2_{{\ts}_1}+ m^2_W )) \nonumber  \\
      & &   + m^2_W ( f^3_2 - f^2_2 ( 3 f_3 + \mgrav^2 )+ (f_3+ \mgrav^2)^2 m^2_{{\ts}_1}
                                         +f_2 ( 2 f^2_3 -  \mgrav^2
					 m^2_{{\ts}_1} +f_3( \mgrav^2+m^2_W) ) ) ] ,\nonumber  \\
					 \\
X_{2pij}&=& -\frac{4\, \sigma_{2ij}}{\mgrav^2 m^2_W} (-f_2+f_3+ \mgrav^2) [  h_6 + h_7 -
f^2_1 \mgrav^2 ] , \\
X_{2kij}&=& -\frac{4\, \sigma_{2ij}}{\mgrav^2 m^2_W} (-f_2+f_3+ \mgrav^2) [  h_6 + h_7 -
f^2_1 \mgrav^2 ] , 
\eeqn
and  $h_6=f_1 f_3 (f_2-f_3)$,  $h_7=  m^2_W ( f_2 (-f_2+f_3)+ \mgrav^2 m^2_{{\ts}_1} )$,
and $\sigma_{1ij}=\Gamma_{1ij} V_i - \Gamma_{2ij} A_i, \, \sigma_{2ij}
=\Gamma_{1ij} V_i + \Gamma_{2ij} A_i$.
We also define $\Gamma_{1j}=\Sigma_{5j} S_i+ \Delta_{5j} P_i$, 
$\Gamma_{2ij}=\Sigma_{5j} P_i + \Delta_{5j} S_i$,
with $\Sigma_{5j}, \, \Delta_{5j}$ defined previously.


\begin{thebibliography}{99}

\bibitem{EHNOS}
J. Ellis, J.S. Hagelin, D.V. Nanopoulos, K.A. Olive
and M. Srednicki, Nucl. Phys. B {\bf 238} (1984) 453; see also
H. Goldberg, Phys. Rev. Lett. {\bf 50} (1983) 1419.

\bibitem{FOS94}
  T.~Falk, K.~A.~Olive and M.~Srednicki,
  Phys.\ Lett.\ B {\bf 339} (1994) 248
  [arXiv:hep-ph/9409270].
  
 \bibitem{vcmssm}
 J.~R.~Ellis, K.~A.~Olive, Y.~Santoso and V.~C.~Spanos,
  Phys.\ Rev.\ D {\bf 70}, 055005 (2004)
  [arXiv:hep-ph/0405110].

\bibitem{gdm}
  J.~R.~Ellis, K.~A.~Olive, Y.~Santoso and V.~C.~Spanos,
  Phys.\ Lett.\ B {\bf 588} (2004) 7
  [arXiv:hep-ph/0312262].

\bibitem{FengGDM}
J.~L.~Feng, A.~Rajaraman and F.~Takayama,
  Phys.\ Rev.\ Lett.\  {\bf 91} (2003) 011302
  [arXiv:hep-ph/0302215];
  Phys.\ Rev.\ D {\bf 68} (2003) 063504
  [arXiv:hep-ph/0306024].

\bibitem{FengGDM2}
  J.~L.~Feng, S.~Su and F.~Takayama,
  Phys.\ Rev.\ D {\bf 70} (2004) 075019
  [arXiv:hep-ph/0404231].

\bibitem{FengGDM3}
  J.~L.~Feng, S.~f.~Su and F.~Takayama,
  Phys.\ Rev.\ D {\bf 70} (2004) 063514
  [arXiv:hep-ph/0404198].





\bibitem{otherGDM2}
K.~Kohri, T.~Moroi and A.~Yotsuyanagi,
  Phys.\ Rev.\ D {\bf 73} (2006) 123511
  [arXiv:hep-ph/0507245].

\bibitem{otherGDM3}
  K.~Jedamzik, K.~Y.~Choi, L.~Roszkowski and R.~Ruiz de Austri,
  JCAP {\bf 0607} (2006) 007
  [arXiv:hep-ph/0512044];
D.~G.~Cerdeno, K.~Y.~Choi, K.~Jedamzik, L.~Roszkowski and R.~Ruiz de Austri,
  JCAP {\bf 0606} (2006) 005
  [arXiv:hep-ph/0509275].
  
  \bibitem{steffan}
F.~D.~Steffen,
  JCAP {\bf 0609}, 001 (2006)
  [arXiv:hep-ph/0605306].


\bibitem{Bench3}
  A.~De Roeck, J.~R.~Ellis, F.~Gianotti, F.~Moortgat, K.~A.~Olive and L.~Pape,
  arXiv:hep-ph/0508198.
  
  \bibitem{Are}
    J.~R.~Ellis, A.~R.~Raklev and O.~K.~Oye,
  JHEP {\bf 0610}, 061 (2006)
  [arXiv:hep-ph/0607261].

\bibitem{Feng+Smith}
  J.~L.~Feng and B.~T.~Smith,
  Phys.\ Rev.\ D {\bf 71} (2005) 015004
  [Erratum-ibid.\ D {\bf 71} (2005) 0109904]
  [arXiv:hep-ph/0409278].


\bibitem{Nojiri}
 K.~Hamaguchi, Y.~Kuno, T.~Nakaya and M.~M.~Nojiri,
  Phys.\ Rev.\ D {\bf 70} (2004) 115007
  [arXiv:hep-ph/0409248].
  
 \bibitem{othercosm}
E.~Holtmann, M.~Kawasaki, K.~Kohri and T.~Moroi,
Phys.\ Rev.\ D {\bf 60}, 023506 (1999)
[arXiv:hep-ph/9805405];
M.~Kawasaki, K.~Kohri and T.~Moroi,
Phys.\ Rev.\ D {\bf 63} (2001) 103502
[arXiv:hep-ph/0012279];
K.~Kohri,
Phys.\ Rev.\ D {\bf 64} (2001) 043515
[arXiv:astro-ph/0103411].

\bibitem{CEFO}
R.~H.~Cyburt, J.~R.~Ellis, B.~D.~Fields and K.~A.~Olive,
  Phys.\ Rev.\ D {\bf 67} (2003) 103521
  [arXiv:astro-ph/0211258].


\bibitem{EOV}
  J.~R.~Ellis, K.~A.~Olive and E.~Vangioni,
  Phys.\ Lett.\ B {\bf 619} (2005) 30
  [arXiv:astro-ph/0503023].

\bibitem{KKM}
M.~Kawasaki, K.~Kohri and T.~Moroi,
  Phys.\ Lett.\ B {\bf 625} (2005) 7
  [arXiv:astro-ph/0402490];
  Phys.\ Rev.\ D {\bf 71} (2005) 083502
  [arXiv:astro-ph/0408426].
      
\bibitem{Maxim}
  M.~Pospelov,
  arXiv:hep-ph/0605215.


\bibitem{otherBoundS}
K.~Kohri and F.~Takayama,
  arXiv:hep-ph/0611016;
M.~Kaplinghat and A.~Rajaraman,
  Phys.\ Rev.\ D {\bf 74} (2006) 103004
  [arXiv:astro-ph/0606209].


\bibitem{CEFOS}
  R.~H.~Cyburt, J.~Ellis, B.~D.~Fields, K.~A.~Olive and V.~C.~Spanos,
  JCAP {\bf 0611} (2006) 014
  [arXiv:astro-ph/0608562].



\bibitem{KKKM}
  T.~Kanzaki, M.~Kawasaki, K.~Kohri and T.~Moroi,
  arXiv:hep-ph/0609246.
  
\bibitem{ER}
J.~R.~Ellis and S.~Rudaz,
  Phys.\ Lett.\ B {\bf 128} (1983) 248.
 
 \bibitem{BDD}
  C.~Boehm, A.~Djouadi and M.~Drees,
  Phys.\ Rev.\ D {\bf 62} (2000) 035012
  [arXiv:hep-ph/9911496].

  
\bibitem{stopco}
  J.~R.~Ellis, K.~A.~Olive and Y.~Santoso,
  Astropart.\ Phys.\  {\bf 18} (2003) 395
  [arXiv:hep-ph/0112113].

\bibitem{Tevatron}
T.~Phillips, talk at DPF 2006, Honolulu, Hawaii, October 2006, \\
{\tt http://www.phys.hawaii.edu/indico/contributionDisplay.py?contribId=454\&amp;\\
sessionId=186\&amp;confId=3}.


\bibitem{luty}
J.~Kang, M.~A.~Luty and S.~Nasri,
  arXiv:hep-ph/0611322.

  
  \bibitem{moroi}
  T.~Moroi,
  arXiv:hep-ph/9503210.

\bibitem{mt171p4}
  E.~Brubaker {\it et al.}  [Tevatron Electroweak Working Group],
  arXiv:hep-ex/0608032.
 
\bibitem{GL}
S.~J.~J.~Gates and O.~Lebedev,
  Phys.\ Lett.\ B {\bf 477} (2000) 216
  [arXiv:hep-ph/9912362].
  
\bibitem{PDG}
W.~M.~Yao {\it et al.}  [Particle Data Group],
  J.\ Phys.\ G {\bf 33} (2006) 1.

\bibitem{sigmab}
CDF Collaboration, {\tt http://www-cdf.fnal.gov/physics/new/bottom/060921.blessed-sigmab/}.
   
\bibitem{ADPRW}
A.~Arvanitaki, S.~Dimopoulos, A.~Pierce, S.~Rajendran and J.~G.~Wacker,
  arXiv:hep-ph/0506242.
  
\bibitem{nunnemann}
  T.~Nunnemann,
  PoS {\bf HEP2005} (2006) 320
  [arXiv:hep-ex/0602039].
  
\bibitem{gallo}
E.~Gallo, talk at ICHEP 2006, Moscow, August 2006, \\
{\tt http://ichep06.jinr.ru/reports/15\_12p10\_Gallo.pdf}.

\bibitem{bsg}
S.~Chen {\it et al.}  [CLEO Collaboration],
Phys.\ Rev.\ Lett.\  {\bf 87} (2001) 251807
[arXiv:hep-ex/0108032];
P.~Koppenburg {\it et al.}  [Belle Collaboration],
Phys.\ Rev.\ Lett.\  {\bf 93} (2004) 061803
[arXiv:hep-ex/0403004].
B.~Aubert {\it et al.}  [BaBar Collaboration],
arXiv:hep-ex/0207076.


\bibitem{bsgth}
 M.~Ciuchini, G.~Degrassi, P.~Gambino and G.~F.~Giudice,
  Nucl.\ Phys.\ B {\bf 527} (1998) 21
  [arXiv:hep-ph/9710335];
  Nucl.\ Phys.\ B {\bf 534} (1998) 3
  [arXiv:hep-ph/9806308];
  C. Degrassi, P. Gambino and G.~F. Giudice,
JHEP {\bf 0012} (2000) 009 [arXiv:hep-ph/0009337];
M.~Carena, D.~Garcia, U.~Nierste and C.~E.~Wagner,
Phys. Lett. B {\bf 499} (2001) 141 
[arXiv:hep-ph/0010003]; 
P.~Gambino and M.~Misiak,
Nucl.\ Phys.\ B {\bf 611} (2001) 338; 
D.~A.~Demir and K.~A.~Olive,
Phys.\ Rev.\ D {\bf 65} (2002) 034007
[arXiv:hep-ph/0107329];
F.~Borzumati, C.~Greub and Y.~Yamada,
  Phys.\ Rev.\ D {\bf 69} (2004) 055005
  [arXiv:hep-ph/0311151];
  T.~Hurth, 
  Rev.\ Mod.\ Phys.\  {\bf 75} (2003) 1159
  [arXiv:hep-ph/0212304].

\bibitem{mh}
LEP Higgs Working Group for Higgs boson searches, OPAL Collaboration,
ALEPH Collaboration, DELPHI Collaboration and L3
Collaboration,
Phys.\ Lett.\ B {\bf 565} (2003) 61 [arXiv:hep-ex/0306033].
{\it Search for neutral Higgs bosons at LEP}, paper submitted to 
ICHEP04, Beijing,
LHWG-NOTE-2004-01, ALEPH-2004-008, DELPHI-2004-042, L3-NOTE-2820,
OPAL-TN-744, \\
{\tt 
http://lephiggs.web.cern.ch/LEPHIGGS/papers/August2004{\_}MSSM/index.html}.


\bibitem{FH} 
  S.~Heinemeyer, W.~Hollik and G.~Weiglein, 
                    {\em Comp. Phys. Comm.} {\bf 124} 2000 76,
                    hep-ph/9812320; 
                    {\em Eur. Phys. J.} {\bf C 9} (1999) 343, 
                    hep-ph/9812472.
                    The codes are accessible via
                    {\tt www.feynhiggs.de} .
                    
 \bibitem{xhiggs}
H.~E.~Haber, R.~Hempfling and A.~H.~Hoang,
  Z.\ Phys.\ C {\bf 75}, 539 (1997)
  [arXiv:hep-ph/9609331];
M.~Carena, H.~E.~Haber, S.~Heinemeyer, W.~Hollik, C.~E.~M.~Wagner and G.~Weiglein,
  Nucl.\ Phys.\ B {\bf 580}, 29 (2000)
  [arXiv:hep-ph/0001002].

 
\bibitem{ourNUHM}
J.~R.~Ellis, K.~A.~Olive and Y.~Santoso,
  Phys.\ Lett.\ B {\bf 539} (2002) 107
  [arXiv:hep-ph/0204192];
J.~R.~Ellis, T.~Falk, K.~A.~Olive and Y.~Santoso,
  Nucl.\ Phys.\ B {\bf 652} (2003) 259
  [arXiv:hep-ph/0210205].
  
\bibitem{baerNUHM}
  H.~Baer, A.~Mustafayev, S.~Profumo, A.~Belyaev and X.~Tata,
  JHEP {\bf 0507} (2005) 065
  [arXiv:hep-ph/0504001].

  
  \bibitem{jittoh}
  T.~Jittoh, J.~Sato, T.~Shimomura and M.~Yamanaka,
  Phys.\ Rev.\ D {\bf 73} (2006) 055009
  [arXiv:hep-ph/0512197].
  
  \bibitem{GS}
  K.~Griest and D.~Seckel,
  Phys.\ Rev.\ D {\bf 43} (1991) 3191.

\bibitem{CEFOS2}
  R.~H.~Cyburt, J.~Ellis, B.~D.~Fields, K.~A.~Olive and V.~C.~Spanos, in
  preparation. 


\end{thebibliography}
\end{document}